\documentclass[aps,prx,10pt,twocolumn,showpacs,showkeys,footinbib,superscriptaddress]{revtex4-2}%superscriptaddress

\usepackage[utf8x]{inputenc}
\usepackage{amsmath}
\usepackage{amssymb}
\usepackage{graphicx}
\usepackage[caption=false]{subfig}
\usepackage{pstricks}
\usepackage{color}
\usepackage{relsize}
\usepackage{bm}
\usepackage{braket}
\usepackage{epstopdf}

\usepackage{hyperref}
\usepackage[all]{hypcap}

\renewcommand{\i}{\ensuremath{\mathrm{i}}}
\newcommand{\e}{\ensuremath{\mathrm{e}}}
\renewcommand{\d}{\ensuremath{\mathrm{d}}}

\begin{document}
\title{Topological Josephson Heat Engine}

\author{Benedikt Scharf}
\email[Corresponding author:]{benedikt.scharf@ur.de}
\affiliation{Institute for Theoretical Physics and Astrophysics and W\"{u}rzburg-Dresden Cluster of Excellence ct.qmat, University of W\"{u}rzburg, Am Hubland, 97074 W\"{u}rzburg, Germany}
\author{Alessandro Braggio}
\affiliation{NEST, Istituto Nanoscienze-CNR and Scuola Normale Superiore, I-56127 Pisa, Italy}
\author{Elia Strambini}
\affiliation{NEST, Istituto Nanoscienze-CNR and Scuola Normale Superiore, I-56127 Pisa, Italy}
\author{Francesco Giazotto}
\affiliation{NEST, Istituto Nanoscienze-CNR and Scuola Normale Superiore, I-56127 Pisa, Italy}
\author{Ewelina M. Hankiewicz}
\affiliation{Institute for Theoretical Physics and Astrophysics and W\"{u}rzburg-Dresden Cluster of Excellence ct.qmat, University of W\"{u}rzburg, Am Hubland, 97074 W\"{u}rzburg, Germany}

\date{\today}

%\pacs{...}
%\keywords{Majorana bound states}

\maketitle

\section{Abstract}
Topological superconductors represent a fruitful playing ground for fundamental research as well as for potential applications in fault-tolerant quantum computing. Especially Josephson junctions based on topological superconductors remain intensely studied both theoretically and experimentally. The characteristic property of these junctions is their $4\pi$-periodic ground-state fermion parity in the superconducting phase difference. Using such topological Josephson junctions, we introduce the concept of a topological Josephson heat engine. We discuss how this engine can be implemented as a Josephson-Stirling cycle in topological superconductors, thereby illustrating the potential of the intriguing and fruitful marriage between topology and coherent thermodynamics. It is shown that the Josephson-Stirling cycle constitutes a highly versatile thermodynamic machine with different modes of operation controlled by the cycle temperatures. Finally, the thermodynamic cycle reflects the hallmark $4\pi$-periodicity of topological Josephson junctions and could therefore be envisioned as a complementary approach to test topological superconductivity.

\begin{figure}[t]
\centering
% trim={<left> <lower> <right> <upper>}
\includegraphics*[trim=0 0 4.2cm 0,width=8.5cm]{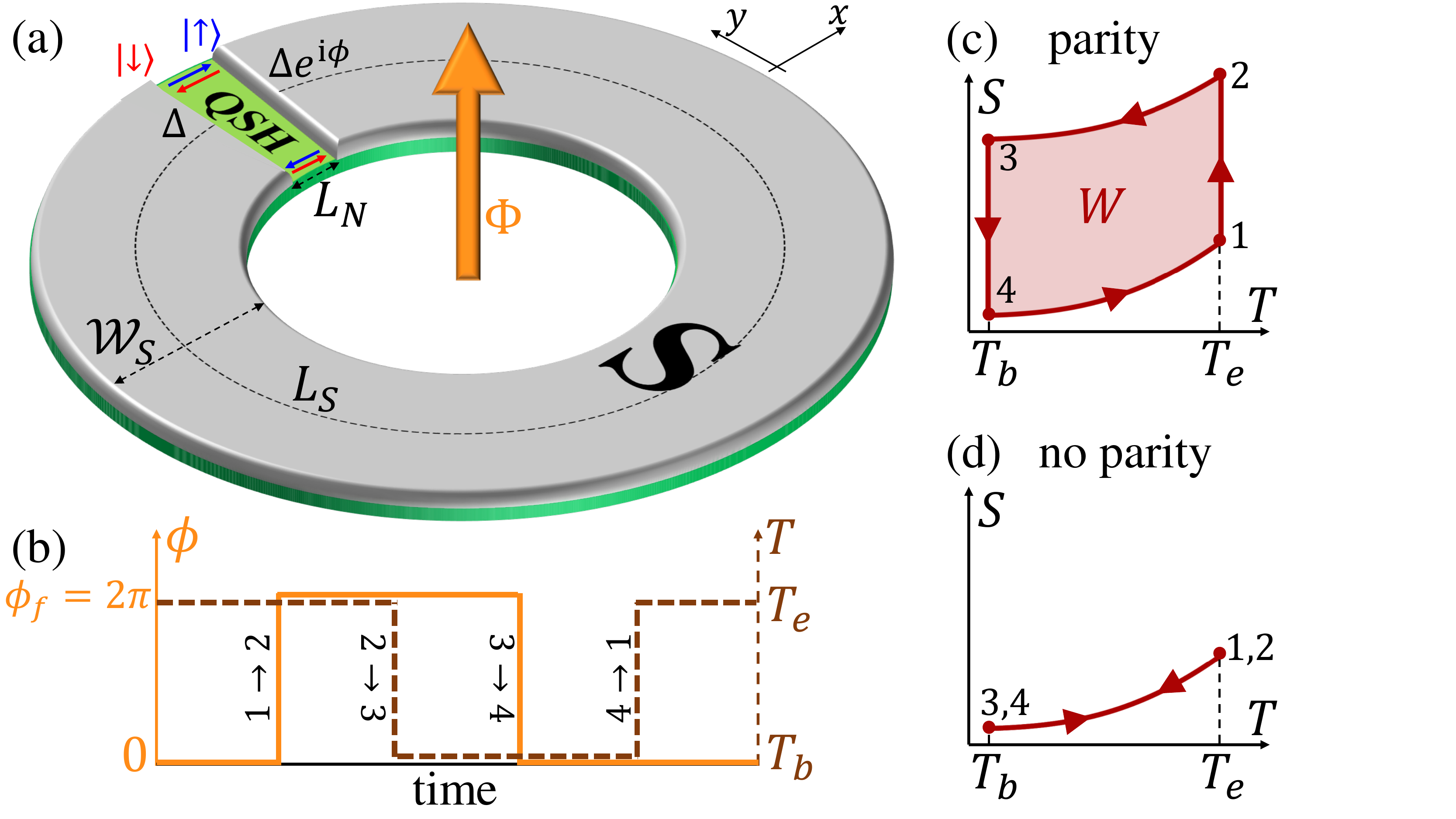}
\caption{Concept of a topological Josephson heat engine. (a) Scheme: A quantum spin Hall (QSH) insulator is partially covered by an $s$-wave superconductor (S), which proximity-induces pairing into the QSH edge states $\ket{\uparrow}$ and $\ket{\downarrow}$, thus defining (proximitized) superconducting regions. A magnetic flux $\Phi$ induces a superconducting phase difference $\phi$ across the normal QSH weak link. The proximitized QSH system is at temperature $T$. (b) Timeline defining the four sequences of $\phi$ and $T$ to implement the Josephson-Stirling thermodynamic cycle: $\bm{1}\equiv(\phi=0,T_\mathrm{e})$, $\bm{2}\equiv(\phi=\phi_\mathrm{f},T_\mathrm{e})$, $\bm{3}\equiv(\phi=\phi_\mathrm{f},T_\mathrm{b})$ and $\bm{4}\equiv(\phi=0,T_\mathrm{b})$. (c,d) Josephson-Stirling cycle in the $(T,S)$ plane for $\phi_\mathrm{f}=2\pi$ with and without parity conservation, respectively. Here, $S$ is the total entropy of the system. The area enclosed by the cycle in the $(T,S)$ plane corresponds to the total heat $Q$ absorbed, which is equivalent to the total work $W$ done during the cycle. In this setup, $W\neq0$ only if parity is conserved due to the trivial $2\pi$-periodicity of the non-topological engine.}\label{fig:scheme}
\end{figure}

\section{Introduction}
The promise of fault-tolerant quantum computing has made topological superconductors the focus of intense research during the past decade~\cite{Hasan2010:RMP,Qi2011:RMP}. In this context, topological Josephson junctions based on nanowires~\cite{Lutchyn2010:PRL,Murthy2019:arxiv} or on topological insulators~\cite{Fu2008:PRL,Houzet2013:PRL,Beenakker2013:PRL,Crepin2014:PRL,Sothmann2017:NJP,Tkachov2015:PRB,Murani2019:PRL} provide an alternative route for probing topological superconductivity. Their topological nature is reflected in a ground-state fermion parity that is $4\pi$-periodic in the superconducting phase difference $\phi$. Finding unambiguous experimental evidence for this $4\pi$-periodicity still proves a difficult task, however~\cite{Rokhinson2012:NP,Wiedenmann2016:NC,Laroche2019:NC,Kayyalha2019:PRL,Deacon2017:PRX}, and it is, therefore, desirable to have several different and complementary approaches. Here we propose a topological Josephson heat engine implemented by a Josephson-Stirling cycle and discuss its thermodynamic~\cite{Giazotto2006:RMP,Giazotto2012:N,Fornieri2016:NN,Fornieri2017:NN,Virtanen2017:PRB,Vischi2019:SR,Vischi2019:E} properties. Using a Josephson junction based on a quantum spin Hall (QSH) insulator as an example, we show how topological Josephson junctions represent versatile thermodynamic machines with various operating modes. Moreover, the thermodynamic cycle properties reflect the $4\pi$-periodicity of the topological ground state, distinguishing between parity-conserving and non-parity-conserving engines. Interestingly, we find that parity conservation generally boosts both the efficiency and power of the topological heat engine with respect to its non-topological counterpart. Our results are applicable not only to QSH-based junctions but also to any topological Josephson junction and establish topological Josephson heat engines as a novel testbed for the $4\pi$-periodicity of the ground-state fermion parity by its entropic signature.

\section{Results and Discussion}
\subsection{Basic Ideas}
In our proposed setup [Fig.~\ref{fig:scheme}(a)], an external magnetic flux controls the superconducting phase bias $\phi$ across the junction. The temperature $T$ of the QSH system is assumed to be externally modulated compared to the bath temperature $T_\mathrm{b}$. For example, this could be done with radiative heating of the system~\cite{Pendry1999:JPCM,Qiu1993:JHT,Baffou2013:LPR} or by having the superconductors acting as reservoirs whose temperature is controlled via resistors or superconductor/insulator/superconductor tunnel junctions~\cite{Giazotto2006:RMP}.

A Josephson-Stirling cycle~\cite{Vischi2019:E} is composed by a sequence of i) an isothermal phase change of $\phi=0\to\phi_\mathrm{f}$ at an externally set temperature $T=T_\mathrm{e}$, followed by ii) an isophasic temperature change $T=T_\mathrm{e}\to T_\mathrm{b}$ at constant $\phi=\phi_\mathrm{f}$, iii) an isothermal phase change of $\phi=\phi_\mathrm{f}\to0$ at $T=T_\mathrm{b}$, and iv) an isophasic temperature change $T=T_\mathrm{b}\to T_\mathrm{e}$ at $\phi=0$ to complete the cycle [Fig.~\ref{fig:scheme}(b)]. If the reference phase $\phi_\mathrm{f}$ is chosen as (an integer multiple of) $\phi_\mathrm{f}=2\pi$, the work released by the engine crucially differs between a setup without fermion-parity constraints and a setup with constant fermion parity. In the former case, the free energy and other thermodynamic quantities are $2\pi$-periodic. This requires that no work or heat is generated or absorbed during each of the isothermal phase changes $\phi=0\to2\pi$ and $\phi=2\pi\to0$. If we assume, on the other hand, that the fermion parity can be kept constant throughout all processes, the thermodynamic quantities are $4\pi$-periodic. Work and heat are then exchanged with the reservoirs during the isothermal phase changes $\phi=0\to2\pi$ and $\phi=2\pi\to0$. Hence, for $\phi_\mathrm{f}=2\pi$ a topological heat engine releases work only when parity can be conserved [Figs.~\ref{fig:scheme}(c,d)].

\subsection{Model and Thermodynamic Properties}
While the concepts outlined above are expected for any topological Josephson junction, we will discuss them explicitly for the example of a short, topological Josephson junction based on a QSH insulator. Here, the pairing in the superconducting (S) regions is induced from nearby $s$-wave superconductors [see Fig.~\ref{fig:scheme}(a), also for the coordinate system]. Assuming two independent edges of the QSH system, the corresponding Bogoliubov-de Gennes (BdG) Hamiltonian for the QSH edge states then reads
\begin{multline}\label{eq:BDGHamSimple}
\hat{H}_{s,\sigma}=\left(s\sigma v_\mathrm{F}\hat{p}_x-\mu_\mathrm{S}\right)\tau_z+V_0L_\mathrm{N}\delta(x)\tau_z\\
+\Delta\left[\tau_x\cos\Phi(x)-\tau_y\sin\Phi(x)\right],
\end{multline}
where $s=\uparrow/\downarrow\equiv\pm1$ describes the natural (out-of-plane) spin projection, $\sigma=t/b\equiv\pm1$ the top and bottom edges, and $\tau_j$ (with $j=x,y,z$) denote Pauli matrices of the particle-hole degrees of freedom (see Supplementary Notes~\ref{App:SimpHam}).

We study a short junction with a normal QSH region of width $L_\mathrm{N}$, approximated by a $\delta$-like profile. The proximity-induced pairing amplitude is $\Delta$ and we use the phase convention $\Phi(x)=\Theta(x)\phi$ to describe the superconducting phase difference $\phi$ between the two S regions. Furthermore, $\hat{p}_x$ denotes the momentum operator, and $V_0$ is the potential difference between the normal QSH and proximitized S regions. We employ a scattering approach to determine the Andreev bound states and the continuum spectrum of Eq.~(\ref{eq:BDGHamSimple}), from which we obtain the free energy---up to some additive $\phi$-independent contributions---as
\begin{equation}\label{eq:FE_Delta}
F_0(\phi,T)=-2k_\mathrm{B}T\ln\left[2\cosh\left(\frac{\Delta\cos\frac{\phi}{2}}{2k_\mathrm{B}T}\right)\right]
\end{equation}
with the Boltzmann constant $k_\mathrm{B}$ and the temperature $T$ of the QSH states (see Supplementary Notes~\ref{App:ABSandCont} and~\ref{App:FE}). Here, Eq.~(\ref{eq:FE_Delta}) arises solely from the Andreev bound state energies and the prefactor 2 takes into account contributions from the top and bottom edges.

Equation~(\ref{eq:FE_Delta}) describes a situation where the states of the system have equilibrium occupations without any external constraints. If fermion-parity conservation is enforced, the free energy acquires an additional term and becomes
\begin{multline}\label{eq:FEp_Delta}
F_p(\phi,T)=\quad\quad\\
-2k_\mathrm{B}T\ln\left[\cosh\left(\frac{\Delta\cos\frac{\phi}{2}}{2k_\mathrm{B}T}\right)+p\e^{J_\mathrm{S}}\sinh\left(\frac{\Delta\cos\frac{\phi}{2}}{2k_\mathrm{B}T}\right)\right],
\end{multline}
where we use the convention that $p=\pm1$ corresponds to even and odd ground-state parity, respectively. In Eq.~(\ref{eq:FEp_Delta}), we again omit additive $\phi$-independent contributions to $F_p$, which are also parity independent. The contribution
\begin{equation}\label{eq:Js}
%J_\mathrm{S}(T)=-\frac{2\Delta^2}{\pi k_\mathrm{B}TE_\mathrm{S}}\int\limits_1^\infty\d\xi\;\frac{\sqrt{\xi^2-1}}{\sinh\left(\xi\Delta/k_\mathrm{B}T\right)}
J_\mathrm{S}(T)=-\frac{2}{\pi k_\mathrm{B}TE_\mathrm{S}}\int\limits_\Delta^\infty\d\epsilon\;\frac{\sqrt{\epsilon^2-\Delta^2}}{\sinh\left(\epsilon/k_\mathrm{B}T\right)}
\end{equation}
originates from the superconducting electrodes, where the energy scale $E_\mathrm{S}=\hbar v_\mathrm{F}/L_\mathrm{S}$ is related to the total length $L_\mathrm{S}$ of the superconducting QSH edge~\cite{Beenakker2013:PRL}. Following Refs.~\cite{Beenakker1992,Beenakker2013:PRL}, we have assumed rigid boundary conditions in writing down Eqs.~(\ref{eq:FE_Delta}) and~(\ref{eq:FEp_Delta}) and do therefore not take into account the inverse proximity effect since $L_\mathrm{S}\gg L_\mathrm{N}$~\cite{Virtanen2017:PRB}.

\begin{figure}[t]
\centering
\includegraphics*[width=8.5cm]{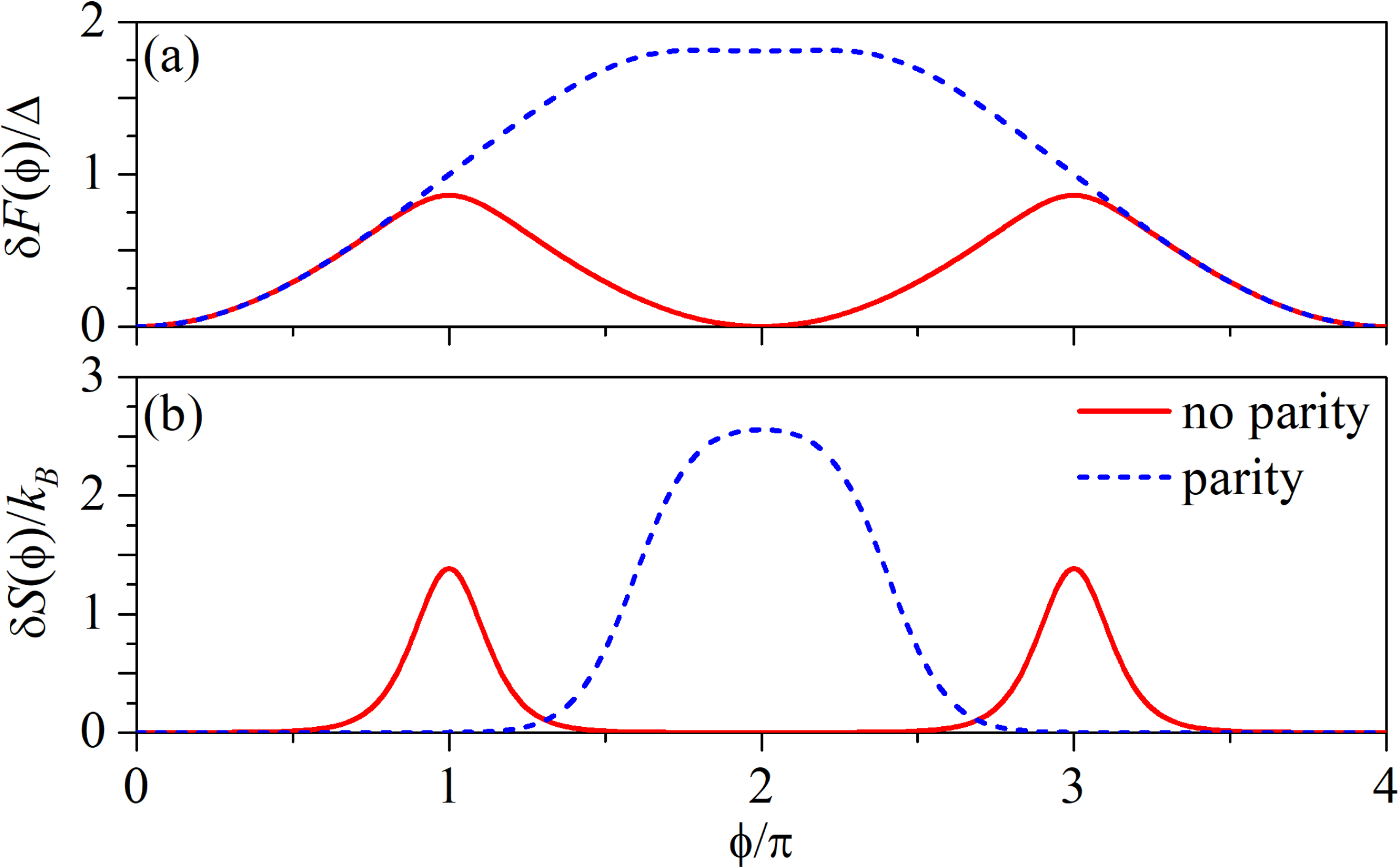}
\caption{Phase-dependent thermodynamic quantities. (a) Phase-dependent variation of the total free energy $\delta F(\phi)=F(\phi)-F(0)$ and (b) entropy $\delta S(\phi)=S(\phi)-S(0)$ of the system for $k_\mathrm{B}T=0.1\Delta$ and $E_\mathrm{S}=0.165\Delta$ without [$F=F_0$ given by Eq.~(\ref{eq:FE_Delta})] and with parity constraints [$F=F_p$ given by Eq.~(\ref{eq:FEp_Delta})]. If parity constraints are enforced, we choose the branch $p=1$. In contrast to the current $I$, the exact values of $F$ and $S$ (as well as $C$) at a given $\phi$ also require knowledge of the $\phi$-independent contributions omitted in Eqs.~(\ref{eq:FE_Delta}) and~(\ref{eq:FEp_Delta}). To overcome this difficulty, we measure $F$ and $S$ with respect to their values at $\phi=0$, thereby canceling the offset due to the $\phi$-independent contributions.}\label{fig:FE}
\end{figure}

The total free energy $F$ of the QSH junction, given by Eqs.~(\ref{eq:FE_Delta}) or~(\ref{eq:FEp_Delta}), allows us to calculate the total Josephson current via~\cite{Beenakker1992}
\begin{equation}\label{eq:JoCur}
I(\phi,T)=\frac{2e}{\hbar}\frac{\partial F(\phi,T)}{\partial\phi},
\end{equation}
where $e$ is the elementary charge, and the entropy via
\begin{equation}\label{eq:Entropy}
S(\phi,T)=-\frac{\partial F(\phi,T)}{\partial T}.
\end{equation}
From $S$, one can subsequently obtain the heat capacity of the junction
\begin{equation}\label{eq:HeatCap}
C(\phi,T)=T\frac{\partial S(\phi,T)}{\partial T}.
\end{equation}
Importantly, Eq.~(\ref{eq:FE_Delta}) is $2\pi$-periodic in $\phi$, while Eq.~(\ref{eq:FEp_Delta}) is $4\pi$-periodic. Consequently, the quantities derived from Eqs.~(\ref{eq:FE_Delta}) or~(\ref{eq:FEp_Delta}) inherit the respective periodicities. This is illustrated by Figs.~\ref{fig:FE}(a,b), which show $F$ and $S$ for junctions without and with parity constraints. For simplicity, we assume a temperature-independent proximity gap, $\Delta(T)\approx\Delta(T=0)$ and $\partial\Delta/\partial T\approx0$, during our calculations, which is reasonably valid for the setup considered here(see Supplementary Notes~\ref{App:FE}).

\subsection{Thermodynamic Processes}
For the Josephson-Stirling cycle, we need to describe different thermodynamic processes. We study quasi-static processes, during which the system passes through quasi-equilibrium states. Then, the work done and heat released during a process $i\to f$ are $W_{i\to f}=-\hbar/(2e)\int\d\phi\; I(\phi,T)$ and $Q_{i\to f}=\int\d S\; T$, respectively. The sign convention is such that $W_{i\to f}$ is positive when the system releases work to the environment, while $Q_{i\to f}$ is positive when the system absorbs heat from the environment.

For an isothermal process where $\phi$ is changed from $\phi_i\to\phi_f$ at constant $T$, $W_{i\to f}=-[F(\phi_f,T)-F(\phi_i,T)]$ and $Q_{i\to f}=T[S(\phi_f,T)-S(\phi_i,T)]$ can be directly obtained from Eqs.~(\ref{eq:FE_Delta}) and~(\ref{eq:FEp_Delta}) and their temperature derivatives. During an isophasic process, $T$ is changed from $T_i\to T_f$ at constant $\phi$. In this case, $W_{i\to f}=0$, while
\begin{equation}\label{eq:Qisophasic}
Q_{i\to f}=\int\limits_{T_i}^{T_f}\d T\;\left[C_0(T)+\delta C(\phi,T)\right]
\end{equation}
can be calculated from the total heat capacity. The $\phi$-dependent contribution $\delta C(\phi,T)=C(\phi,T)-C_0(T)$ can be directly calculated from Eqs.~(\ref{eq:FE_Delta}) or~(\ref{eq:FEp_Delta}) and its derivatives and is measured with respect to $\phi=0$. In principle, we also need to account for the $\phi$-independent contribution $C_0(T)$ arising from the terms omitted in Eqs.~(\ref{eq:FE_Delta}) and~(\ref{eq:FEp_Delta}). For additional details, we refer to Supplementary Notes~\ref{App:TDproc} and~\ref{App:Stirling}, where $C_0(T)$ is calculated using the BCS DOS.

\begin{figure}[t]
\centering
\includegraphics*[width=8.5cm]{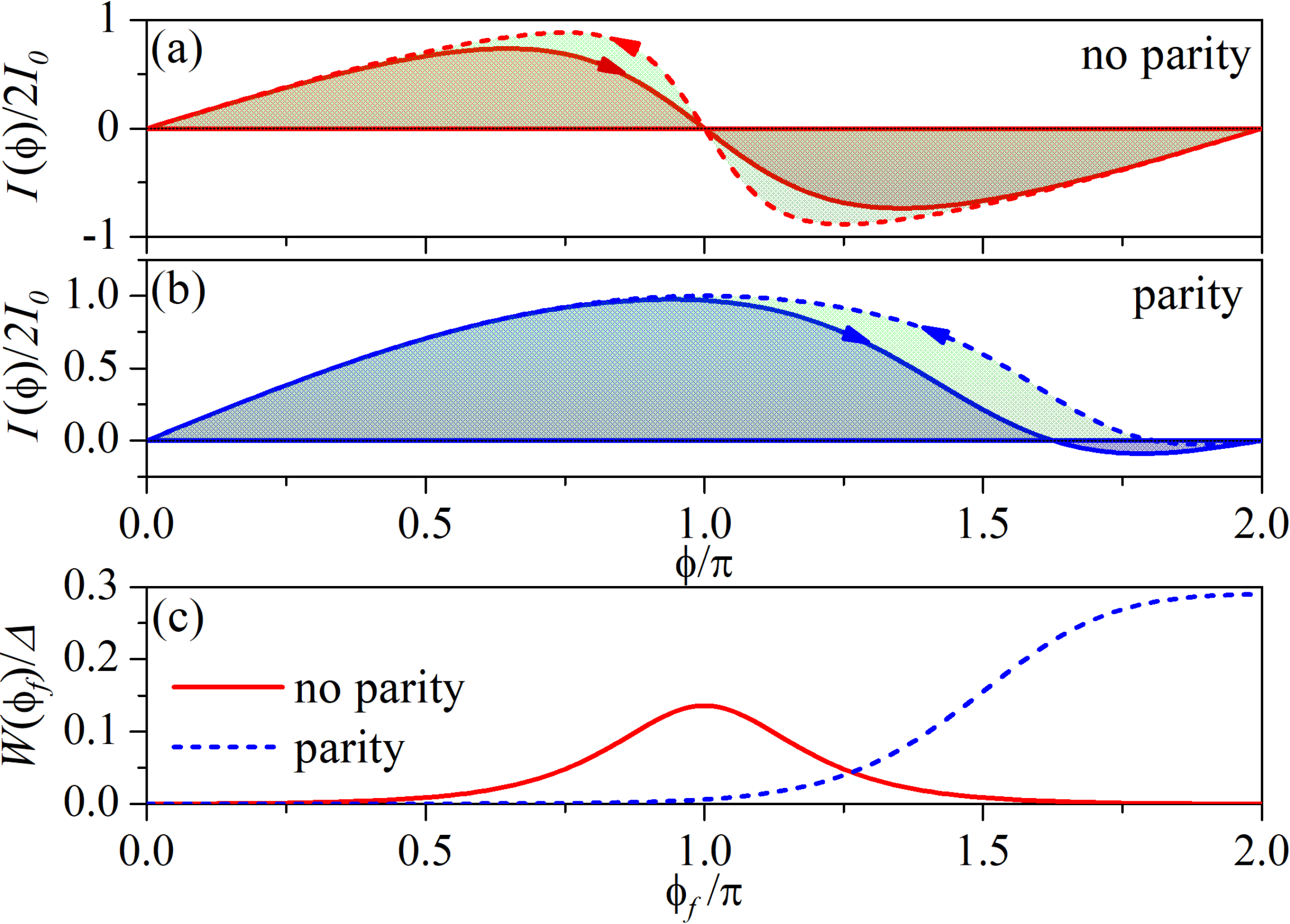}
\caption{Current-phase relation and work. (a,b) Isothermal current-phase relation at $k_\mathrm{B}T=0.1\Delta$ (dashed lines) and $k_\mathrm{B}T=0.2\Delta$ (solid lines) (a) without and (b) with parity constraints. Here, the shaded areas between each curve and the $\phi$ axis correspond to the work performed during an isothermal change $\phi=0\to2\pi$. The areas above and below the $\phi$ axis compensate each other in~(a) for an isothermal change $\phi=0\to2\pi$ and no work is released. In panels~(a,b), the arrows indicate the direction of the changes in $\phi$ for a Josephson-Stirling engine with $k_\mathrm{B}T_\mathrm{e}=0.2\Delta$, $k_\mathrm{B}T_\mathrm{b}=0.1\Delta$, and $\phi_\mathrm{f}=2\pi$. The total work $W$ performed by these engines is represented by the shaded green areas between the dashed and solid curves. (c) Total work $W$ released by a Josephson-Stirling engine with $k_\mathrm{B}T_\mathrm{e}=0.2\Delta$ and $k_\mathrm{B}T_\mathrm{b}=0.1\Delta$ as a function of the maximal phase change $\phi_\mathrm{f}$ during the cycle. In all panels, $E_\mathrm{S}=0.165\Delta$.}\label{fig:StirlingWork}
\end{figure}

\subsection{Josephson-Stirling Cycle}
We are now in a position to explicitly compute the total work and heat produced during each of the processes of the Josephson-Stirling cycle introduced above [Fig.~\ref{fig:scheme}(b)]. As mentioned above, $W_{1\to2}$ and $W_{3\to4}$ correspond to integrals over the current-phase relation, but can also be computed directly from $F$. The total work $W=W_{1\to2}+W_{3\to4}$ of each cycle thus coincides with the difference between the integrated areas over the current-phase relation [Figs.~\ref{fig:StirlingWork}(a,b)]. The heat exchanged with the hot ($T=T_\mathrm{e}$) and cold reservoirs ($T=T_\mathrm{b}$) is $Q_\mathrm{e}=Q_{1\to2}+Q_{4\to1}$ and $Q_\mathrm{b}=Q_{2\to3}+Q_{3\to4}$, respectively. Conservation of energy dictates $W=Q$, where $Q=Q_\mathrm{e}+Q_\mathrm{b}$ is the total heat exchange during the cycle. Note that in our setup, there are no separate hot and cold reservoirs, but the environment acts successively as hot and cold reservoir.

In Fig.~\ref{fig:StirlingWork}(c), we show $W$ as a function of the reference phase $\phi_\mathrm{f}$ and compare the case without and with parity constraints. Without parity conservation, $W$ is maximal for $\phi_\mathrm{f}=\pi$, whereas $W=0$ for $\phi_\mathrm{f}=2\pi$. The latter is a consequence of the $2\pi$-periodicity of $F_0(\phi,T)=F_0(\phi+2\pi,T)$, causing $W_{1\to2}=-[F_0(\phi_\mathrm{f},T_\mathrm{e})-F_0(0,T_\mathrm{e})]$ and $W_{3\to4}=-[F_0(0,T_\mathrm{b})-F_0(\phi_\mathrm{f},T_\mathrm{b})]$ to each vanish for $\phi_\mathrm{f}=2\pi$. If fermion parity is kept constant, on the other hand, $F_p(\phi,T)=F_p(\phi+4\pi,T)$ and $W$ is maximal for $\phi_\mathrm{f}=2\pi$. A topological heat engine with $\phi_\mathrm{f}=2\pi$ thus releases work only if parity is conserved and can thus serve as a test for the $4\pi$-periodicity of the ground-state fermion parity.

\begin{figure}[t]
\centering
\includegraphics*[width=8.5cm]{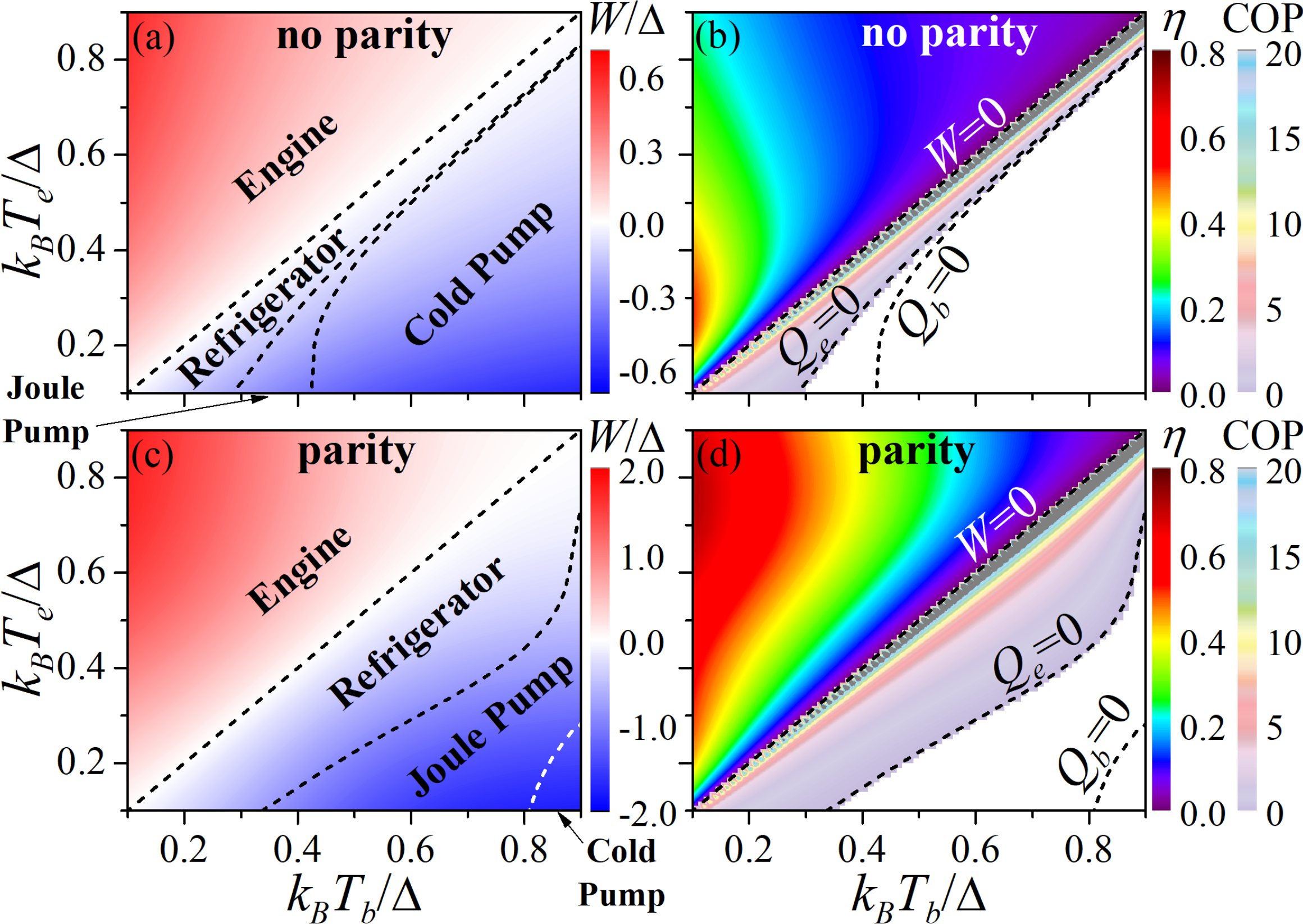}
\caption{Josephson-Stirling cycle. Total work $W$ and efficiency $\eta$ or Coefficient of Performance (COP) as functions of the reservoir temperatures $T_\mathrm{e}$ and $T_\mathrm{b}$ (a,b) without and (c,d) with parity constraints. In both cases, $E_\mathrm{S}=0.165\Delta$. The different operating modes of the cycle are indicated in panels~(a,c): For refrigerators, the cycle absorbs heat $Q_\mathrm{e}$ from the cooled subsystem with temperature $T_\mathrm{e}$ and releases heat $Q_\mathrm{b}<0$ with $|Q_\mathrm{b}|>Q_\mathrm{e}$ to the heat sink with temperature $T_\mathrm{b}$. If $T_\mathrm{e}<T_\mathrm{b}$, $Q_\mathrm{e}<0$, and $Q_\mathrm{b}<0$, the machine is a Joule pump that completely converts work into heat released to the reservoirs. On the other hand, if $T_\mathrm{e}<T_\mathrm{b}$, $Q_\mathrm{e}<0$, and $W<0$, while $Q_\mathrm{b}>0$, the machine acts as a cold pump transferring heat from the hot ($T=T_\mathrm{b}$) to the cold reservoir ($T=T_\mathrm{e}$).}\label{fig:Stirling}
\end{figure}

\subsection{Different Operating Modes}
Until now, we have focused only on an engine. Depending on the relative values of $T_\mathrm{e}$ and $T_\mathrm{b}$, the Josephson-Stirling cycle can, however, exhibit also other operating modes. This is illustrated by Fig.~\ref{fig:Stirling}, which shows $W$ and cycle efficiencies for different combinations of $T_\mathrm{e}$ and $T_\mathrm{b}$ without [Figs.~\ref{fig:Stirling}(a,b)] and with [Figs.~\ref{fig:Stirling}(c,d)] parity constraints. Here, $\phi_\mathrm{f}$ is chosen to yield the maximal work, that is, $\phi_\mathrm{f}=\pi$ without parity constraints [Fig.~\ref{fig:Stirling}(a)] and $\phi_\mathrm{f}=2\pi$ if parity is conserved [Fig.~\ref{fig:Stirling}(c)].

For $T_\mathrm{e}>T_\mathrm{b}$, the Josephson-Stirling cycle/machine acts as an engine: The machine absorbs $Q_\mathrm{e}>0$ from the hot reservoir and releases $|Q_\mathrm{b}|<Q_\mathrm{e}$ to the cold reservoir. Hence, $W>0$ is done on the environment and the engine efficiency is given by $\eta=W/Q_\mathrm{e}$. A comparison of the engine efficiency and maximal power shows that a parity-conserving engine is on average more efficient and more powerful than its non-parity-conserving implementation [Figs.~\ref{fig:Stirling}(b,d)]. We interpret the stronger power as due to an increased phase space available: To obtain a finite work, the work integral can be integrated over a $0-2\pi$ range if parity is preserved, whereas one needs to remain within the $0-\pi$ range without parity conservation. Secondly, the lower efficiency of the non-parity-conserving engine can be understood as due to the competition between mutually exclusive processes with opposite parities. Indeed, Eq.~(\ref{eq:FEp_Delta}) shows that the additional parity-related terms contribute with different signs to $F_p$, implying opposite contributions to the heat exchange. Consequently, the non-parity-preserving engine can be interpreted as a thermal machine composed of two mutually exclusive engines working in an opposite manner, thereby reducing the total efficiency.

If $T_\mathrm{e}<T_\mathrm{b}$, the systems with and without parity conservation act as refrigerators with a Coefficient of Performance $\mathrm{COP}=Q_\mathrm{e}/|W|$~\cite{Vischi2019:E} or as Joule or cold pumps. Controlling $T_\mathrm{e}$ vs $T_\mathrm{b}$ thus enables multiple operating modes of the Josephson-Stirling cycle. If the cycle is set up as in Fig.~\ref{fig:scheme}(b), refrigerators as well as Joule and cold pumps require that $T_\mathrm{e}<T_\mathrm{b}$. While it is possible by superconductor/insulator/superconductor cooling to bring $T_\mathrm{e}$ below $T_\mathrm{b}$~\cite{Giazotto2006:RMP}, a more promising way to realize refrigerators, Joule or cold pumps is to shift the cycle by interchanging the initial and final phases, $\phi=0$ and $\phi=\phi_\mathrm{f}$. Such a setup implies the same phase diagrams as in Fig.~\ref{fig:Stirling} but with $T_\mathrm{e}$ and $T_\mathrm{b}$ interchanged (see Supplementary Notes~\ref{App:PlotsStirling}). Hence, the 'shifted' Josephson-Stirling cycle can be used to realize operating modes other than engines, making it a highly versatile thermodynamic machine.

\subsection{Potential Routes to Experimental Realization}
Importantly, a topological Josephson heat engine implemented as a Josephson-Stirling cycle can be used to test the hallmark $4\pi$-periodicity of the phase-dependent ground-state fermion parity. In this implementation, a major challenge is to fastly modulate the temperature of the proximitized QSH junction while preserving its fermion parity. While this condition precludes galvanic channels of heat transfer to the QSH system, others such as phononic~\cite{Li2012:RMP}, photonic~\cite{Bosisio2016:PRB,Ronzani2018:NP} or radiative~\cite{Pendry1999:JPCM} channels could be used. In such setups, the timescale of the temperature modulation could be as low as $0.1$ ns (see Supplementary Notes~\ref{App:QP}), much smaller than typical quasiparticle poisoning timescales, which are of the order of 1 $\mu$s~\cite{Rainis2012:PRB,Virtanen2013:PRB,Frombach2020:PRB}. Hence, implementing a parity conserving Josephson-Stirling cycle appears feasible. The work per cycle can be experimentally determined by measuring the current-phase relation during each isothermal process to compute the work integrals. Such current-phase measurements have, for example, been successfully performed in topological junctions with scanning superconducting quantum interference device microscopes~\cite{Sochnikov2015:PRL,Hart2019:PRB} (see also Supplementary Notes~\ref{App:Measurement}).

\subsection{Conclusions}
In this manuscript, we have explicitly discussed topological Josephson heat engines for the example of a short QSH-based Josephson junction. Since the concept is only based on the $4\pi$-periodicity of the ground-state fermion parity, it is also applicable to long as well as nanowire-based topological Josephson junctions. Schemes to detect signatures of the $4\pi$-periodic ground-state parity in topological Josephson junctions often require careful interpretation of the measurements~\cite{Dominguez2017:PRB,Chiu2019:PRB}. It is therefore desirable to have several different, complementary approaches. As their properties reflect the $4\pi$-periodic ground-state of topological Josephson junctions, topological Josephson heat engines could be used as such a complementary setup to test the topological superconductivity as long as fermion parity can be conserved. However, even without conserving the ground-state fermion parity, topological Josephson junctions represent versatile machines with various operating modes.

\vspace{3mm}

\textbf{Acknowledgments:} B.S. and E.M.H. acknowledge funding by the Deutsche Forschungsgemeinschaft (DFG, German Research Foundation) through SFB 1170, Project-ID 258499086, through Grant No. HA 5893/4-1 within SPP 1666 and through the W\"urzburg-Dresden Cluster of Excellence on Complexity and Topology in Quantum Matter -- ct.qmat (EXC 2147, Project-ID 390858490) as well as by the ENB Graduate School on Topological Insulators. E.S., A.B. and F.G acknowledge partial financial support from the EU’s Horizon 2020 research and innovation program under Grant Agreement No. 800923 (SUPERTED), the CNR-CONICET cooperation program "Energy conversion in quantum nanoscale hybrid devices", the SNS-WIS jointlab QUANTRA funded by the Italian Ministry of Foreign Affairs and International Cooperation and the Royal Society through the International Exchanges between the UK and Italy (Grants No. IES R3 170054 and IEC R2192166). This publication was supported by the Open Access Publication Fund of the University of W\"{u}rzburg.

\vspace{1mm}

\textbf{Author contributions:} E.H. and F.G. conceived the idea of applying the concepts of heat engines to topological Josephson junctions. These concepts were then further developed and refined into the theoretical formulation presented in this work after intense discussions with B.S., A.B. and E.S.. B.S. performed the calculations. All authors contributed to the writing of the manuscript.

\vspace{1mm}

\textbf{Additional information:} Supplementary Information is available in the online version of the paper.

\vspace{1mm}

\textbf{Competing financial interests:} The Authors declare no Competing Financial or Non-Financial Interests.

\vspace{1mm}

\textbf{Data Availability:} The data that support the findings of this study are available within the paper and its Supplementary Information. Additional data are available from the corresponding author upon request.

%%%%%%%%%% Prefix a "S" to all equations, figures, tables and reset the counter %%%%%%%%%%
\setcounter{equation}{0}
\setcounter{figure}{0}
\setcounter{table}{0}
\setcounter{section}{0}
\makeatletter
\renewcommand{\theequation}{S\arabic{equation}}
\renewcommand{\thefigure}{S\arabic{figure}}
\renewcommand{\thesection}{S\arabic{section}}
%%%%%%%%%%%%%%%%%%%%%%%%%%%%%%%%%%%%%%%%%%%%%%%%%%%%%%%%%%%%%%%%%%%%%%%%%%%%%%%%%%%%%%%%%%

%\appendix
\begin{widetext}

\section{Simplified Hamiltonian}\label{App:SimpHam}
Here, we provide additional details on the derivation of the simplified BdG Hamiltonian~(1) in the main text, which serves as the starting point of our calculations. We assume that the two edges of the QSH system are separated by a distance $\mathcal{W}_S$ large enough such that there is no overlap between states from opposite edges. With the basis order $\left(\hat{\psi}_{t,\uparrow},\hat{\psi}_{t,\downarrow},\hat{\psi}_{b,\uparrow},\hat{\psi}_{b,\downarrow},\hat{\psi}^\dagger_{t,\downarrow},-\hat{\psi}^\dagger_{t,\uparrow},\hat{\psi}^\dagger_{b,\downarrow},-\hat{\psi}^\dagger_{b,\uparrow}\right)$, the BdG Hamiltonian describing the Josephson junction is then given by
\begin{equation}\label{eq:BDGHam}
\hat{H}_\mathrm{BdG}=\left(v_\mathrm{F}\hat{p}_xs_z\sigma_z-\mu_\mathrm{S}\right)\tau_z+V_0h(x)\tau_z\\
+\Delta(x)\left[\tau_x\cos\Phi(x)-\tau_y\sin\Phi(x)\right].
\end{equation}
Here, $s_j$, $\sigma_j$ and $\tau_j$ (with $j=x,y,z$) denote Pauli matrices describing spin, top/bottom edge and particle-hole degrees of freedom, respectively. Note that unit matrices are not written explicitly in Eq.~(\ref{eq:BDGHam}). The N region is described by the profile $h(x)$ and the potential $V_0$, which can also be viewed as describing the difference between the chemical potentials in the S and N regions, $\mu_\mathrm{S}$ and $\mu_\mathrm{N}=\mu_\mathrm{S}-V_0$. Here, we use a $\delta$-barrier model with $h(x)=L_\mathrm{N}\delta(x)$ and $\Delta(x)=\Delta$. A more general approach to the junction would be to use a finite N region with $h(x)=\Theta(L_\mathrm{N}/2-|x|)$ and $\Delta(x)=\Delta\Theta(|x|-L_\mathrm{N}/2)$. Such a model is capable of describing both a short as well as a long junction. As explained in Supplementary Notes~\ref{App:ABSandCont} below, the $\delta$-barrier model is still suitable to capture the essential physics of a short junction based on QSH edge states.

Since $\left[\hat{H}_\mathrm{BdG},s_z\right]=\left[\hat{H}_\mathrm{BdG},\sigma_z\right]=0$, the wave functions can be described by the good quantum numbers $s=\uparrow/\downarrow\equiv\pm1$ for the spin projection in $z$ direction and $\sigma=\mathrm{t}/\mathrm{b}\equiv\pm1$ for the top and bottom edges. Hence, an eigenstate $\Psi_{s,\sigma}(x)$ of the BdG Hamiltonian~(\ref{eq:BDGHam}) can be written as $\Psi_{s,\sigma}(x)=\psi_{s,\sigma}(x)\otimes\eta_\sigma\otimes\chi_s$, which satisfies
\begin{equation}\label{eq:BDG}
\hat{H}_{s,\sigma}\psi_{s,\sigma}(x)=E\psi_{s,\sigma}(x),
\end{equation}
where
\begin{equation}\label{eq:BDGHamSimple0}
\hat{H}_{s,\sigma}=\left(s\sigma v_\mathrm{F}\hat{p}_x-\mu_\mathrm{S}\right)\tau_z+V_0L_\mathrm{N}\delta(x)\tau_z\\
+\Delta\left[\tau_x\cos\Phi(x)-\tau_y\sin\Phi(x)\right],
\end{equation}
[see also Eq.~(1) in the main text]. The spinors $\chi_s$ and $\eta_\sigma$ are the eigenvectors of the Pauli matrices $s_z$ and $\sigma_z$, respectively. They satisfy $s_z\chi_s=s\chi_s$ and $\sigma_z\eta_\sigma=\sigma\eta_\sigma$.

\section{Computing the Andreev and continuum spectra}\label{App:ABSandCont}
In order to compute the ABS as well as the $\phi$-dependent part of the continuum DOS $\rho_c(\epsilon,\phi)$, we employ a wave-function matching approach to solve Eq.~(\ref{eq:BDG}).

\subsection{Andreev bound states}\label{App:ABS}
For ABS, that is, states with $|\epsilon|<\Delta$ and thus states localized around the N region, we can make the piecewise ansatz
\begin{equation}\label{eq:ABSansatz}
\psi_{s,\sigma}(x)=\frac{1}{\sqrt{L_{tot}}}\left\{\begin{array}{l}
a_{s,\sigma}\left(\begin{array}{l}1\\C_{s\sigma}\end{array}\right)\e^{\i s\sigma k_\mathrm{F}x}\e^{\kappa_\epsilon x},\; x<0,\\
\quad\\
\quad\\
e_{s,\sigma}\left(\begin{array}{l}1\\0\end{array}\right)\e^{\i s\sigma(k_\mathrm{N}+k_\epsilon)x}+h_{s,\sigma}\left(\begin{array}{l}0\\1\end{array}\right)\e^{\i s\sigma(k_\mathrm{N}-k_\epsilon)x},\; 0<x<L_\mathrm{N},
\quad\\
\quad\\
b_{s,\sigma}\left(\begin{array}{l}1\\C_{-s\sigma}\e^{-\i\phi}\end{array}\right)\e^{\i s\sigma k_\mathrm{F}x}\e^{-\kappa_\epsilon x},\; x>L_\mathrm{N}
\end{array}\right.
\end{equation}
for a junction with a finite N region. The ansatz for a $\delta$-junction is similar to Eq.~(\ref{eq:ABSansatz}), with the states $\psi_{s,\sigma}(x<0)$ given by the first line of Eq.~(\ref{eq:ABSansatz}) and $\psi_{s,\sigma}(x>0)$ given by the third line of Eq.~(\ref{eq:ABSansatz}). In Eq.~(\ref{eq:ABSansatz}), $C_\pm=(\epsilon\pm\i\sqrt{\Delta^2-\epsilon^2})/\Delta$, $k_\mathrm{F}=\mu_\mathrm{S}/(\hbar v_\mathrm{F})$, $\kappa_\epsilon=\sqrt{\Delta^2-\epsilon^2}/(\hbar v_\mathrm{F})$, $k_\mathrm{N}=\mu_\mathrm{N}/(\hbar v_\mathrm{F})$, $k_\epsilon=\epsilon/(\hbar v_\mathrm{F})$, and $L_{tot}=L_\mathrm{N}+L_\mathrm{S}$ is the unit length of the entire S/N/S edge. The ansatz~(\ref{eq:ABSansatz}) for $\psi_{s,\sigma}(x)$ has been chosen in such a way that Eq.~(\ref{eq:BDG}) is satisfied in each S and N region separately and that $\lim\limits_{x\to\pm\infty}\psi_{s,\sigma}(x)=0$. The coefficients $a_{s,\sigma}$, $b_{s,\sigma}$, $e_{s,\sigma}$, and $h_{s,\sigma}$ have to be determined from the boundary conditions at the S/N interfaces.

For the $\delta$-barrier, the boundary condition can be obtained by integrating Eq.~(\ref{eq:BDG}) from $x=-\eta$ to $x=\eta$ with $\eta\to0^+$. The corresponding procedure~\cite{MatosAbiague2003:PRA,Sothmann2016:PRB,Scharf2016:PRL} yields
\begin{equation}\label{eq:BCdelta}
\psi_{s,\sigma}(0^+)=\e^{-\i s\sigma Z_0}\psi_{s,\sigma}(0^-),
\end{equation}
where $Z_0=V_0L_\mathrm{N}/(\hbar v_\mathrm{F})$. For a finite barrier, on the other hand, the boundary conditions require $\psi_{s,\sigma}(x)$ to be continuous at the S/N interfaces,
\begin{equation}\label{eq:BCfinite}
\psi_{s,\sigma}(0^+)=\psi_{s,\sigma}(0^-),\quad\psi_{s,\sigma}(L_\mathrm{N}^+)=\psi_{s,\sigma}(L_\mathrm{N}^-).
\end{equation}

First, we consider the $\delta$-barrier model, for which we invoke the boundary condition~(\ref{eq:BCdelta}) and require a nontrivial solution for the coefficients $a_{s,\sigma}$ and $b_{s,\sigma}$. This procedure yields a transcendental equation for the ABS energy $\epsilon$ in the form of
\begin{equation}\label{eq:transABSdelta}
\arccos\left(\frac{\epsilon}{\Delta}\right)=-s\sigma\frac{\phi}{2}+\pi n,
\end{equation}
where $n$ is an integer that has to be chosen in such a way that $0\leq\arccos\left(\epsilon/\Delta\right)\leq\pi$. Solving Eq.~(\ref{eq:transABSdelta}), we obtain the ABS
\begin{equation}\label{eq:delta_ABS}
\epsilon^\sigma_s(\phi)=-s\sigma\mathrm{sgn}\left(\sin\frac{\phi}{2}\right)\Delta\cos\frac{\phi}{2},
\end{equation}
describing two ABS $s=\uparrow/\downarrow$ per edge ($\sigma=\mathrm{t}/\mathrm{b}$). At a given edge, these two ABS $s=\uparrow/\downarrow$ exhibit a protected crossing at $\phi=\pi$, which also corresponds to a change in the ground-state fermion parity~\cite{Ioselevich2011:PRL,Beenakker2013:PRL}. Equation~(\ref{eq:delta_ABS}) is used to compute the free energies in the main text (see Supplementary Notes~\ref{App:FE} below). 

For a finite N region, we use the boundary conditions~(\ref{eq:BCfinite}) and require a nontrivial solution for the coefficients $a_{s,\sigma}$, $b_{s,\sigma}$, $e_{s,\sigma}$, and $h_{s,\sigma}$. Now, we obtain the transcendental equation
\begin{equation}\label{eq:transABSfinite}
\arccos\left(\frac{\epsilon}{\Delta}\right)-\frac{\epsilon L_\mathrm{N}}{\hbar v_\mathrm{F}}=-s\sigma\frac{\phi}{2}+\pi n,
\end{equation}
where $n$ is again an integer chosen in such a way that $0\leq\arccos\left(\epsilon/\Delta\right)\leq\pi$. The second term on the left-hand side of Eq.~(\ref{eq:transABSfinite}) takes into account the finite width of the N region. This term becomes only important in long junctions, $\Delta\gg\hbar v_\mathrm{F}/L_\mathrm{N}$, where it causes multiple subbands to appear in the Andreev spectrum. These additional bound states in long junctions cannot be captured with a $\delta$-model, which always yields two ABS per edge. Still for short junctions, $\Delta\ll\hbar v_\mathrm{F}/L_\mathrm{N}$, Eq.~(\ref{eq:transABSfinite}) tends to Eq.~(\ref{eq:transABSdelta}).

\subsection{Continuum states}\label{App:Continuum}
Above, we have determined the discrete Andreev spectrum. If we now turn to the continuous spectrum with $\epsilon>\Delta$, we have to modify the ansatz~(\ref{eq:ABSansatz}) to account for propagating wave functions. An incident electron-like quasiparticle propagating from $x\to-\infty$ to $x\to\infty$ with $\epsilon>\Delta$ can be described by the ansatz
\begin{equation}\label{eq:conansatz}
\psi_{s,\sigma}(x)=\frac{1}{\sqrt{L_{tot}}}\left\{\begin{array}{l}
\left(\begin{array}{l}u\\v\end{array}\right)\e^{\i(k_\mathrm{F}+q_\epsilon)x}+r_\mathrm{eh}^{s,\sigma}\left(\begin{array}{l}v\\u\end{array}\right)\e^{\i(k_\mathrm{F}-q_\epsilon)x},\; x<0,\\
\quad\\
\quad\\
e_{s,\sigma}\left(\begin{array}{l}1\\0\end{array}\right)\e^{\i(k_\mathrm{N}+k_\epsilon)x}+h_{s,\sigma}\left(\begin{array}{l}0\\1\end{array}\right)\e^{\i(k_\mathrm{N}-k_\epsilon)x},\; 0<x<L_\mathrm{N},
\quad\\
\quad\\
t_\mathrm{ee}^{s,\sigma}\left(\begin{array}{l}u\\v\e^{-\i\phi}\end{array}\right)\e^{\i(k_\mathrm{F}+q_\epsilon)x},\; x>L_\mathrm{N},
\end{array}\right.
\end{equation}
if a finite N region is considered. The ansatz for a $\delta$-junction is again similar to Eq.~(\ref{eq:conansatz}), but with $\psi_{s,\sigma}(x<0)$ given by the first line of Eq.~(\ref{eq:conansatz}) and $\psi_{s,\sigma}(x>0)$ given by the third line of Eq.~(\ref{eq:conansatz}). In order for the quasiparticle to be a right mover, $s\sigma=1$ in Eq.~(\ref{eq:conansatz}), that is, $s=\uparrow$ at the top edge and $s=\downarrow$ at the bottom edge. Corresponding equations can be set up for hole-like quasiparticles propagating to the right as well as for quasiparticles propagating to the left. In Eq.~(\ref{eq:conansatz}), $u^2=(1+\sqrt{\epsilon^2-\Delta^2}/\epsilon)/2=1-v^2$, $q_\epsilon=\sqrt{\epsilon^2-\Delta^2}/(\hbar v_\mathrm{F})$, and all other quantities are the same as defined in Eq.~(\ref{eq:ABSansatz}). Note that no additional normalization factors are needed in Eq.~(\ref{eq:conansatz}) because each individual state in the S regions carries the same absolute value of quasiparticle current.

By invoking the boundary conditions~(\ref{eq:transABSdelta}) or~(\ref{eq:transABSfinite}), one can then obtain the reflection and transmission coefficients $r_\mathrm{eh}^{s,\sigma}$ and $t_\mathrm{ee}^{s,\sigma}$ in Eq.~(\ref{eq:conansatz}). For hole-like incident quasiparticles, one can likewise obtain $r_\mathrm{he}^{s,\sigma}$ and $t_\mathrm{hh}^{s,\sigma}$. These states then allow us to set up the $S$ matrix as
\begin{equation}\label{eq:Smat}
S_\mathrm{SNS}=\left(\begin{array}{cc} S_\mathrm{SNS}^\mathrm{t} & 0 \\ 0 & S_\mathrm{SNS}^\mathrm{b} \end{array}\right)
\end{equation}
with
\begin{equation}\label{eq:Smat_t}
S_\mathrm{SNS}^\mathrm{t}=\left(\begin{array}{cccc} t_\mathrm{ee}^{\uparrow,\mathrm{t}} & 0 & 0 & r_\mathrm{he}^{\uparrow,\mathrm{t}} \\ 0 & t_\mathrm{hh}^{\downarrow,\mathrm{t}} & r_\mathrm{eh}^{\downarrow,\mathrm{t}} & 0 \\ 0 & r_\mathrm{he}^{\downarrow,\mathrm{t}} & t_\mathrm{ee}^{\downarrow,\mathrm{t}} & 0 \\ r_\mathrm{eh}^{\uparrow,\mathrm{t}} & 0 & 0 & t_\mathrm{hh}^{\uparrow,\mathrm{t}} \end{array}\right)
\end{equation}
and
\begin{equation}\label{eq:Smat_b}
S_\mathrm{SNS}^\mathrm{b}=\left(\begin{array}{cccc} t_\mathrm{ee}^{\downarrow,\mathrm{b}} & 0 & 0 & r_\mathrm{he}^{\downarrow,\mathrm{b}} \\ 0 & t_\mathrm{hh}^{\uparrow,\mathrm{b}} & r_\mathrm{eh}^{\uparrow,\mathrm{b}} & 0 \\ 0 & r_\mathrm{he}^{\uparrow,\mathrm{b}} & t_\mathrm{ee}^{\uparrow,\mathrm{b}} & 0 \\ r_\mathrm{eh}^{\downarrow,\mathrm{b}} & 0 & 0 & t_\mathrm{hh}^{\downarrow,\mathrm{b}} \end{array}\right).
\end{equation}

The scattering matrix $S^\sigma_\mathrm{SNS}(\epsilon,\phi)$ allows us then to compute the ($\phi$-dependent part of the) continuum DOS of the Josephson junction as
\begin{equation}\label{eq:DOSphase}
\rho^\sigma_0(\epsilon,\phi)=\frac{1}{2\pi\i}\frac{\partial}{\partial\epsilon}\ln\left[\det\left(S^\sigma_\mathrm{SNS}\right)\right].
\end{equation}
Since $S_\mathrm{SNS}$ is block-diagonal in the top/bottom-edge degree of freedom, $\mathrm{det}\left(S_\mathrm{SNS}\right)=\mathrm{det}\left(S_\mathrm{SNS}^\mathrm{t}\right)\mathrm{det}\left(S_\mathrm{SNS}^\mathrm{b}\right)$ can be decomposed in a contribution from the top edge ($\sigma=1$) and one from the bottom edge ($\sigma=-1$). Then, we obtain
\begin{equation}\label{eq:DetSmat_t_delta}
\mathrm{det}\left(S_\mathrm{SNS}^\mathrm{t}\right)=\frac{\epsilon^2-\Delta^2\cos^2(\phi/2)}{\epsilon^2\cos\phi-\Delta^2\cos^2(\phi/2)+\i\epsilon\sqrt{\epsilon^2-\Delta^2}\sin\phi}\frac{\epsilon^2-\Delta^2\cos^2(\phi/2)}{\epsilon^2\cos\phi-\Delta^2\cos^2(\phi/2)-\i\epsilon\sqrt{\epsilon^2-\Delta^2}\sin\phi}=1
\end{equation}
for the top edge of a $\delta$-barrier model if $\epsilon>\Delta$. Note that because spin $s=\uparrow/\downarrow$ is a good quantum number, Eq.~(\ref{eq:DetSmat_t_delta}) also factorizes into separate contributions from $s=\uparrow/\downarrow$. These two factors have opposite complex phases which cancel with each other to give $\mathrm{det}\left(S_\mathrm{SNS}^\mathrm{t}\right)=1$. The contribution from the bottom edge can be obtained by switching $\uparrow\,\leftrightarrow\,\downarrow$. Hence, $\mathrm{det}\left(S_\mathrm{SNS}^\mathrm{b}\right)=\mathrm{det}\left(S_\mathrm{SNS}^\mathrm{t}\right)=1$. Inserting Eq.~(\ref{eq:DetSmat_t_delta}) into Eq.~(\ref{eq:DOSphase}) then yields $\rho^\sigma_0(\epsilon,\phi)=0$. For the $\delta$-model, there is consequently no $\phi$-dependent contribution to the free energy from the continuum states.

If we consider a finite N region, the above discussion still applies, but now
\begin{eqnarray}\label{eq:DetSmat_t_finite}
\mathrm{det}\left(S_\mathrm{SNS}^\mathrm{t}\right)=\frac{\e^{-2\i q_\epsilon L_\mathrm{N}}\left[\epsilon^2-\Delta^2\cos^2\left(\frac{\phi-2k_\epsilon L_\mathrm{N}}{2}\right)\right]}{\epsilon^2\cos\left(\phi-2k_\epsilon L_\mathrm{N}\right)-\Delta^2\cos^2\left(\frac{\phi-2k_\epsilon L_\mathrm{N}}{2}\right)+\i\epsilon\sqrt{\epsilon^2-\Delta^2}\sin\left(\phi-2k_\epsilon L_\mathrm{N}\right)}\nonumber\\
\times\frac{\e^{-2\i q_\epsilon L_\mathrm{N}}\left[\epsilon^2-\Delta^2\cos^2\left(\frac{\phi+2k_\epsilon L_\mathrm{N}}{2}\right)\right]}{\epsilon^2\cos\left(\phi+2k_\epsilon L_\mathrm{N}\right)-\Delta^2\cos^2\left(\frac{\phi+2k_\epsilon L_\mathrm{N}}{2}\right)-\i\epsilon\sqrt{\epsilon^2-\Delta^2}\sin\left(\phi+2k_\epsilon L_\mathrm{N}\right)}
\end{eqnarray}
for the top edge if $\epsilon>\Delta$. The contribution from the bottom edge can be obtained from $\uparrow\,\leftrightarrow\,\downarrow$ and thus again as $\mathrm{det}\left(S_\mathrm{SNS}^\mathrm{t}\right)=\mathrm{det}\left(S_\mathrm{SNS}^\mathrm{b}\right)$. For $L_\mathrm{N}\to0$, $\mathrm{det}\left(S_\mathrm{SNS}^\mathrm{t}\right)=\mathrm{det}\left(S_\mathrm{SNS}^\mathrm{b}\right)\to1$ as in the $\delta$-model. We thus conclude that also for short finite junctions, the continuum states do not significantly affect the $\phi$ dependence of the free energy.

\section{Free energy of a single edge}\label{App:FE}
From the ABS and continuum spectra of Eq.~(\ref{eq:BDG}), one can calculate the free energy $F_\sigma(\phi,T)$ at a single edge $\sigma=\mathrm{t}/\mathrm{b}$. These free energies can in turn be used to obtain the Josephson current and other thermodynamic quantities of a given edge.

\subsection{No parity constraints}\label{App:FENoParity}
Without parity constraints, the free energy is---up to an additive $\phi$-independent contribution---given by~\cite{Beenakker1992,Beenakker2013:PRL}
\begin{equation}\label{eq:FE}
F^\sigma_0(\phi,T)=-k_\mathrm{B}T\left\{\sum\limits_{n,\epsilon^\sigma_n\geq0}\ln\left[2\cosh\left(\frac{\epsilon^\sigma_n(\phi)}{2k_\mathrm{B}T}\right)\right]+\int\limits_0^\infty\d\epsilon\rho^\sigma_c(\epsilon,\phi)\ln\left[2\cosh\left(\frac{\epsilon}{2k_\mathrm{B}T}\right)\right]\right\},
\end{equation}
where $k_\mathrm{B}$ is the Boltzmann constant and $T$ the temperature. The sum over $n$ describes the contribution from the ABS with discrete energies $\epsilon^\sigma_n(\phi)$, while the integral describes the contribution from the continuum states with DOS $\rho^\sigma_c(\epsilon,\phi)$. At this point, it is important to note that due to the lack of spin degeneracy the degeneracy factor is just $g=1$ in Eq.~(\ref{eq:FE})~\cite{Beenakker2013:PRL}. Whereas we can directly insert the Andreev spectrum $\epsilon^\sigma_n(\phi)$ in Eq.~(\ref{eq:FE}), we have to compute the continuum DOS
\begin{equation}\label{eq:DOS}
\rho^\sigma_c(\epsilon,\phi)=\rho^\sigma_0(\epsilon,\phi)+\rho_S(\epsilon).
\end{equation}
The $\phi$-dependent continuous spectrum of the Josephson junction is described by $\rho^\sigma_0(\epsilon,\phi)$, which we compute from the scattering matrix via Eq.~(\ref{eq:DOSphase}) and which vanishes for the $\delta$-model, $\rho^\sigma_0(\epsilon,\phi)=0$, consistent with the expectation that only ABS exhibit a significant $\phi$ dependence in short junctions~\cite{Beenakker1992}.

Following Ref.~\cite{Beenakker2013:PRL}, we also include in Eq.~(\ref{eq:DOS}) a $\phi$-independent term originating from the superconducting electrodes,
\begin{equation}\label{eq:DOSSv2}
\rho_S(\epsilon)=\frac{2}{\pi E_\mathrm{S}}\frac{|\epsilon|\Theta\left(\epsilon^2-\Delta^2\right)}{\sqrt{\epsilon^2-\Delta^2}}.
\end{equation}
Here, the energy scale $E_\mathrm{S}=\hbar v_\mathrm{F}/L_\mathrm{S}$ is determined from the length $L_\mathrm{S}$ of the superconducting electrodes. Inserting the ABS $\epsilon^\sigma_n$ from Eq.~(\ref{eq:delta_ABS}) into Eq.~(\ref{eq:FE}) and omitting all $\phi$-independent terms, which arise from $\rho^\sigma_c(\epsilon,\phi)=\rho_S(\epsilon)$, we arrive at Eq.~(2) in the main text. Hence, Eq.~(2) in the main text describes only the $\phi$-dependent part of the free energy. While $\rho_S(\epsilon)$ is independent of $\phi$ and terms arising from $\rho^\sigma_c(\epsilon,\phi)=\rho_S(\epsilon)$ in Eq.~(\ref{eq:FE}) do therefore not affect the Josephson current in the absence of parity constraints~\cite{Beenakker1992}, $\rho_S(\epsilon)$ can play an important role if the parity is kept constant, as will be explained below.

\subsection{Parity constraints}\label{App:FEParity}
If the fermion parity $p=\pm$ is conserved at a given edge $\sigma=\mathrm{t}/\mathrm{b}$, the free energy acquires an additional contribution due to the parity constraint and reads~\cite{Ioselevich2011:PRL,Beenakker2013:PRL}
\begin{equation}\label{eq:FEp}
F^\sigma_p(\phi,T)=F^\sigma_0(\phi,T)-k_\mathrm{B}T\ln\left\{\frac{1}{2}\left\{1+pP(\phi)\left[\prod\limits_{n,\epsilon^\sigma_n\geq0}\tanh\left(\frac{\epsilon^\sigma_n(\phi)}{2k_\mathrm{B}T}\right)\right]\exp\left[\int\limits_0^\infty\d\epsilon\rho^\sigma_c(\epsilon,\phi)\ln\left(\tanh\left(\frac{\epsilon}{2k_\mathrm{B}T}\right)\right)\right]\right\}\right\}.
\end{equation}
Here, $p=\pm1$ and the function
\begin{equation}\label{eq:DefParity}
P(\phi)=\mathrm{sgn}\left(\cos\frac{\phi}{2}\right)
\end{equation}
describes the ground-state fermion parity as $\phi$ is tuned. From the form of Eq.~(\ref{eq:FEp}), it is clear that even $\phi$-independent terms in $\rho^\sigma_c$ are important in determining the $\phi$ dependence of $F^\sigma_p$. This is why, for example, the contribution from the superconducting electrodes can also affect other quantities such as the Josephson current. If we split the integral over $\rho^\sigma_c$ in Eq.~(\ref{eq:FEp}) into an integral over $\rho^\sigma_0$ and $\rho_S$, define
\begin{equation}\label{eq:JsDef}
J_\mathrm{S}(T)=\int\limits_0^\infty\d\epsilon\rho_S(\epsilon)\ln\left[\tanh\left(\frac{\epsilon}{2k_\mathrm{B}T}\right)\right]=-\frac{2}{\pi k_\mathrm{B}TE_\mathrm{S}}\int\limits_\Delta^\infty\d\epsilon\;\frac{\sqrt{\epsilon^2-\Delta^2}}{\sinh\left(\epsilon/k_\mathrm{B}T\right)}=-\frac{2\Delta^2}{\pi k_\mathrm{B}TE_\mathrm{S}}\int\limits_1^\infty\d\xi\;\frac{\sqrt{\xi^2-1}}{\sinh\left(\xi\Delta/k_\mathrm{B}T\right)},
\end{equation}
and insert the ABS for $\epsilon^\sigma_n$, we arrive at Eq.~(3) in the main text.

For simplicity, we assume a temperature-independent proximity gap, $\Delta(T)\approx\Delta(T=0)$ and $\partial\Delta/\partial T\approx0$, which is reasonably valid if $T\ll T_c$, where $T_c$ is the critical temperature of the parent superconductor. For HgTe-based Josephson junctions on Nb superconductors with a native superconducting gap $\Delta_{Nb}\approx1$ meV~\cite{Oostinga2013:PRX}, the proximity-induced gap in HgTe is $\Delta\approx0.15\Delta_{Nb}=150$ $\mu$eV~\cite{Oostinga2013:PRX,Sochnikov2015:PRL}. The critical temperature of the parent Nb superconductor can therefore be estimated as $T_c\approx0.57\Delta_{Nb}/k_\mathrm{B}\approx6.6$ K. In this manuscript, we focus mainly on operating temperatures corresponding to $k_\mathrm{B}T<0.5\Delta$ ($T<0.87$ K). These temperatures are well below $T_c$ of Nb, justifying our approximation $\partial\Delta/\partial T\approx0$. In this case, $J_\mathrm{S}(T)$ can be approximated by $J_\mathrm{S}(T)\approx-(4\Delta/\pi E_\mathrm{S})K_1(\Delta/k_\mathrm{B}T)$ for $k_\mathrm{B}T\ll\Delta$, where $K_1$ is the modified Bessel function of the second kind. Equations~(\ref{eq:FEp}) and~(\ref{eq:JsDef}) show that even $\phi$-independent contributions from the continuum modify the $\phi$ dependence of the free energy via the factor $\exp(J_\mathrm{S})$ if fermion parity is conserved. This is different compared to short junctions without fermion-parity conservation, where $\phi$-independent continuum contributions do not affect the $\phi$ dependence of the free energy.

Note that, similar to Eq.~(\ref{eq:FE}), Eq.~(\ref{eq:FEp}) yields the same value for the top and bottom edges, $\sigma=\mathrm{t}/\mathrm{b}$. Hence, the total free energy of both edges is simply twice the free energy of a single edge. This applies to the cases without and with fermion-parity conservation.

Finally, we note that we neglect the inductance associated with the superconducting ring here, that is, we assume that both the geometrical and kinetic inductances associated with the ring are negligible. There is, however, no equilibrium current circulating in the superconducting ring and hence no additional contribution to $F$ for integer multiples of $\phi=2\pi$. Without constraints on the fermion parity, there is likewise no current and no inductance contribution for integer multiples of $\phi=\pi$. Similar to Ref.~\cite{Vischi2019:SR}, the value of the ring inductance does therefore not affect the main conclusions of this manuscript, which focuses on the reference phases $\phi=0$ and $\phi=\phi_\mathrm{f}=2\pi$ with parity conservation and on the reference phases $\phi=0$ and $\phi=\phi_\mathrm{f}=\pi$ without parity constraints.

\section{Thermodynamic processes and full phase dependence of the heat capacity}\label{App:TDproc}
For the quasi-static processes studied here, the work and heat released during a process $i\to f$ are $W_{i\to f}=-\hbar/(2e)\int\d\phi\; I(\phi,T)$ and $Q_{i\to f}=\int\d S\; T$, respectively. The sign convention is such that $W_{i\to f}$ is positive when the system releases work to the environment, while $Q_{i\to f}$ is positive when the system absorbs heat from the environment. Throughout the manuscript, we assume $\Delta(T)\approx\Delta(T=0)$ and $\partial\Delta/\partial T\approx0$.

During an isothermal process, $\phi$ is changed from $\phi_i\to\phi_f$ at constant $T$. In this case, the work $W_{i\to f}=-[F(\phi_f,T)-F(\phi_i,T)]$ and the heat exchange $Q_{i\to f}=T[S(\phi_f,T)-S(\phi_i,T)]$ can be directly obtained from Eqs.~(2) and~(3) in the main text and their temperature derivatives. Note that, while $W_{i\to f}$ corresponds to an integral over the current-phase relation, it can also be computed directly from $F$. This relation holds because we do not include the inverse proximity effect in our model.%\footnote{Throughout the paper, we use $E_\mathrm{S}=0.165\Delta$, which for $\Delta=100$ $\mu$eV and $v_\mathrm{F}=5\times10^5$ m/s corresponds to $L_\mathrm{S}=20$ $\mu$m or a radius of around $3.2$ $\mu$m}

For an isophasic process, where $T$ is changed from $T_i\to T_f$ at constant $\phi$, $W_{i\to f}=0$. The heat exchange, on the other hand, is given by
\begin{equation}\label{eq:Qisophasic0}
Q_{i\to f}=\int\limits_{T_i}^{T_f}\d T\;\left[C_0(T)+\delta C(\phi,T)\right]
\end{equation}
and can be calculated from the total heat capacity [see also Eq.~(8) in the main text]. This requires knowledge of the full $T$ dependence of the total heat capacity $C$. Equations~(2)-(7) in the main text can, however, only provide $\delta C(\phi,T)=C(\phi,T)-C(0,T)$ because we omitted additive $\phi$-independent contributions in Eqs.~(2) and~(3). To also include these contributions, we have to calculate the reference capacity $C_0(T)\equiv C(\phi=0,T)$ independently. For $k_\mathrm{B}T\ll\Delta$, we can approximate~\cite{Vischi2019:E}
\begin{equation}\label{eq:Capprox}
\frac{C_0(T)}{\Delta/(\pi E_\mathrm{S})}\approx2k_\mathrm{B}\sqrt{2\pi}\left(\frac{\Delta}{k_\mathrm{B}T}\right)^{3/2}\exp\left(-\frac{\Delta}{k_\mathrm{B}T}\right),
\end{equation}
using the BCS DOS. From $C(\phi,T)=C_0(T)+\delta C(\phi,T)$ and Eq.~(\ref{eq:Qisophasic0}) [equivalent to Eq.~(8) in the main text], one can then compute the heat exchange during an isophasic process.

For the complete Josephson-Stirling cycle, the dependence on the exact form of $C_0(T)$ is cancelled because the initial and final temperatures are interchanged during the two isophasic processes. Consequently, the total heat exchanged during the cycle does not depend on the specific form of Eq.~(\ref{eq:Capprox}) and must be equal to the total work done during the cycle as dictated by conservation of energy. Instead, the choice in Eq.~(\ref{eq:Capprox}) affects the engine efficiency which is determined by the heat exchanged with the reservoir during the processes where the temperature of the QSH edges is changed (see Supplementary Notes~\ref{App:Stirling}).

\section{Josephson-Stirling cycle without parity constraints: Analytics}\label{App:Stirling}
An appealing property of a Josephson-Stirling cycle without parity constraints is that it allows for closed analytical expressions for the work and heat flows. For example, the heat flows during the isophasic processes can then be computed from Eqs.~(6)-(8) in the main text and Eq.~(\ref{eq:Capprox}) above as
\begin{multline}\label{eq:QisophasicNP}
Q_{i\to f}=2\left\{\frac{\Delta}{2}\left[\tanh\left(\frac{\Delta}{k_\mathrm{B}T_f}\right)-\tanh\left(\frac{\Delta}{k_\mathrm{B}T_i}\right)\right]-\frac{\sqrt{2}\Delta^2}{E_\mathrm{S}}\left[\mathrm{Erf}\left(\sqrt{\frac{\Delta}{k_\mathrm{B}T_f}}\right)-\mathrm{Erf}\left(\sqrt{\frac{\Delta}{k_\mathrm{B}T_i}}\right)\right]\right.\\
\left.-\frac{\Delta\cos\phi/2}{2}\left[\tanh\left(\frac{\Delta\cos\phi/2}{k_\mathrm{B}T_f}\right)-\tanh\left(\frac{\Delta\cos\phi/2}{k_\mathrm{B}T_i}\right)\right]\right\},
\end{multline}
where $\mathrm{Erf}$ denotes the error function and the factor 2 accounts for the top and bottom edges.

\begin{figure}[t]
\centering
\includegraphics*[width=12cm]{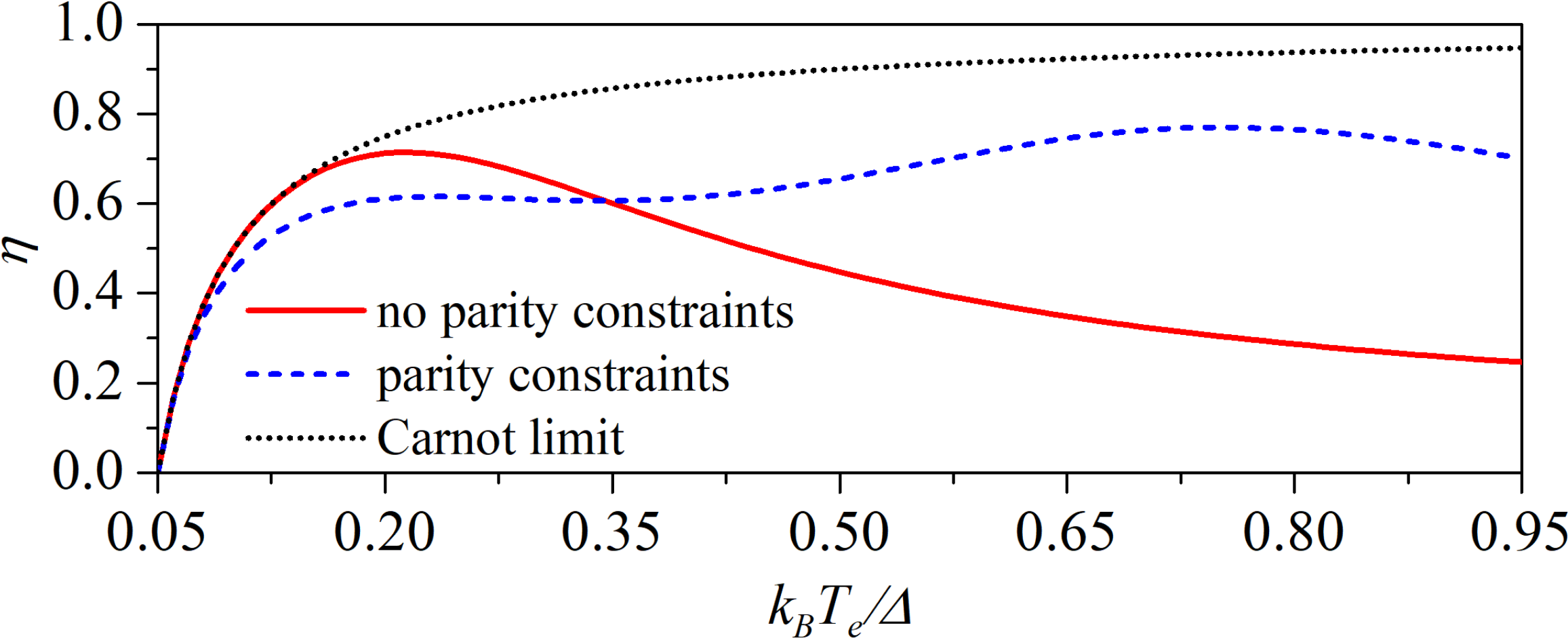}
\caption{Comparison between the efficiency $\eta$ given by Eq.~(\ref{eq:EffNP}) for a non-parity-conserving engine and the numerically calculated efficiency of a parity-conserving engine. Here, $k_\mathrm{B}T_\mathrm{b}=0.05\Delta$ and $E_\mathrm{S}=0.165\Delta$. The black dotted line depicts the Carnot limit $\eta_C=1-T_\mathrm{e}/T_\mathrm{b}$.}\label{fig:StirlingEff}
\end{figure}

For the Josephson-Stirling cycle discussed in the main text with $\phi_\mathrm{f}=\pi$ and without parity constraints, we can calculate the total work $W$ as well as the heat exchanges $Q_\mathrm{e}$ and $Q_\mathrm{b}$. Using Eq.~(\ref{eq:QisophasicNP}) for the isophasic processes as well as $Q_{i\to f}=T[S(\phi_\mathrm{f},T)-S(\phi_\mathrm{i},T)]$ for the isothermal processes, the heat exchange with the reservoir at temperature $T_\mathrm{e}$, $Q_\mathrm{e}=Q_{1\to2}+Q_{4\to1}$, can then be computed as
\begin{equation}\label{eq:QRNP}
Q_\mathrm{e}=2\left\{\frac{\sqrt{2}\Delta^2}{E_\mathrm{S}}\left[\mathrm{Erf}\left(\sqrt{\frac{\Delta}{k_\mathrm{B}T_\mathrm{b}}}\right)-\mathrm{Erf}\left(\sqrt{\frac{\Delta}{k_\mathrm{B}T_\mathrm{e}}}\right)\right]-k_\mathrm{B}T_\mathrm{e}\ln\left[\cosh\left(\frac{\Delta}{2k_\mathrm{B}T_\mathrm{e}}\right)\right]+\frac{\Delta}{2}\tanh\left(\frac{\Delta}{2k_\mathrm{B}T_\mathrm{e}}\right)\right\}.
\end{equation}
Likewise, the heat exchange with the reservoir at temperature $T_\mathrm{b}$ is $Q_\mathrm{b}=Q_{2\to3}+Q_{3\to4}$ and given by
\begin{equation}\label{eq:QLNP}
Q_\mathrm{b}=2\left\{\frac{\sqrt{2}\Delta^2}{E_\mathrm{S}}\left[\mathrm{Erf}\left(\sqrt{\frac{\Delta}{k_\mathrm{B}T_\mathrm{e}}}\right)-\mathrm{Erf}\left(\sqrt{\frac{\Delta}{k_\mathrm{B}T_\mathrm{b}}}\right)\right]+k_\mathrm{B}T_\mathrm{b}\ln\left[\cosh\left(\frac{\Delta}{2k_\mathrm{B}T_\mathrm{b}}\right)\right]-\frac{\Delta}{2}\tanh\left(\frac{\Delta}{2k_\mathrm{B}T_\mathrm{e}}\right)\right\}.
\end{equation}
The total work $W=W_{1\to2}+W_{3\to4}$ during one cycle originates only from the isothermal processes, where $W_{i\to f}=-[F(\phi_\mathrm{f},T)-F(\phi_\mathrm{i},T)]$, which in turn yields
\begin{equation}\label{eq:WNP}
W=2\left\{k_\mathrm{B}T_\mathrm{b}\ln\left[\cosh\left(\frac{\Delta}{2k_\mathrm{B}T_\mathrm{b}}\right)\right]-k_\mathrm{B}T_\mathrm{e}\ln\left[\cosh\left(\frac{\Delta}{2k_\mathrm{B}T_\mathrm{e}}\right)\right]\right\}
\end{equation}
Again, the factors of 2 in Eqs.~(\ref{eq:QRNP})-(\ref{eq:WNP}) are due to considering both the top and bottom edges.

For $T_\mathrm{e}>T_\mathrm{b}$, the machine acts as an engine with a hot reservoir of temperature $T_\mathrm{e}$ and a cold reservoir of temperature $T_\mathrm{b}$. In this case, the efficiency $\eta=W/Q_\mathrm{e}$ of the cycle is
\begin{equation}\label{eq:EffNP}
\eta=\frac{k_\mathrm{B}T_\mathrm{b}\ln\left[\cosh\left(\frac{\Delta}{2k_\mathrm{B}T_\mathrm{b}}\right)\right]-k_\mathrm{B}T_\mathrm{e}\ln\left[\cosh\left(\frac{\Delta}{2k_\mathrm{B}T_\mathrm{e}}\right)\right]}{\frac{\sqrt{2}\Delta^2}{E_\mathrm{S}}\left[\mathrm{Erf}\left(\sqrt{\frac{\Delta}{k_\mathrm{B}T_\mathrm{b}}}\right)-\mathrm{Erf}\left(\sqrt{\frac{\Delta}{k_\mathrm{B}T_\mathrm{e}}}\right)\right]-k_\mathrm{B}T_\mathrm{e}\ln\left[\cosh\left(\frac{\Delta}{2k_\mathrm{B}T_\mathrm{e}}\right)\right]+\frac{\Delta}{2}\tanh\left(\frac{\Delta}{2k_\mathrm{B}T_\mathrm{e}}\right)},
\end{equation}
which is in turn bounded by the Carnot limit $\eta_C=1-T_\mathrm{b}/T_\mathrm{e}$. In the refrigerator mode, the superconductor acts as a cooled subsystem of temperature $T_S=T_\mathrm{e}$ and as a heat sink of temperature $T_S=T_\mathrm{b}$ with $T_\mathrm{e}<T_\mathrm{b}$, $Q_\mathrm{b}<0$ and $0<Q_\mathrm{e}<|Q_\mathrm{b}|$. Then, the coefficient of performance $\mathrm{COP}=Q_\mathrm{e}/|W|$ is given by
\begin{equation}\label{eq:COPNP}
\mathrm{COP}=\frac{\frac{\sqrt{2}\Delta^2}{E_\mathrm{S}}\left[\mathrm{Erf}\left(\sqrt{\frac{\Delta}{k_\mathrm{B}T_\mathrm{b}}}\right)-\mathrm{Erf}\left(\sqrt{\frac{\Delta}{k_\mathrm{B}T_\mathrm{e}}}\right)\right]-k_\mathrm{B}T_\mathrm{e}\ln\left[\cosh\left(\frac{\Delta}{2k_\mathrm{B}T_\mathrm{e}}\right)\right]+\frac{\Delta}{2}\tanh\left(\frac{\Delta}{2k_\mathrm{B}T_\mathrm{e}}\right)}{k_\mathrm{B}T_\mathrm{e}\ln\left[\cosh\left(\frac{\Delta}{2k_\mathrm{B}T_\mathrm{e}}\right)\right]-k_\mathrm{B}T_\mathrm{b}\ln\left[\cosh\left(\frac{\Delta}{2k_\mathrm{B}T_\mathrm{b}}\right)\right]}.
\end{equation}

For illustration, Supplementary Fig.~\ref{fig:StirlingEff} shows the $T_\mathrm{e}$ dependence of the efficiency $\eta$ given by Eq.~(\ref{eq:EffNP}) for fixed $k_\mathrm{B}T_\mathrm{b}=0.05\Delta$ and compares it with the numerically calculated efficiency of a parity-conserving engine. While for small $T_\mathrm{e}$, both efficiencies are close to the Carnot limit $\eta_C$, the engine with parity conservation is much more efficient at higher $T_\mathrm{e}$. This is in agreement with our considerations in the main text concerning a generically higher efficiency of the topological engine with preserved fermion parity.

\renewcommand\thesubfigure{\Alph{subfigure}}
\begin{figure}[t]
\centering
\subfloat[Josephson-Stirling cycle without parity constraints]{\includegraphics[width=8.5cm]{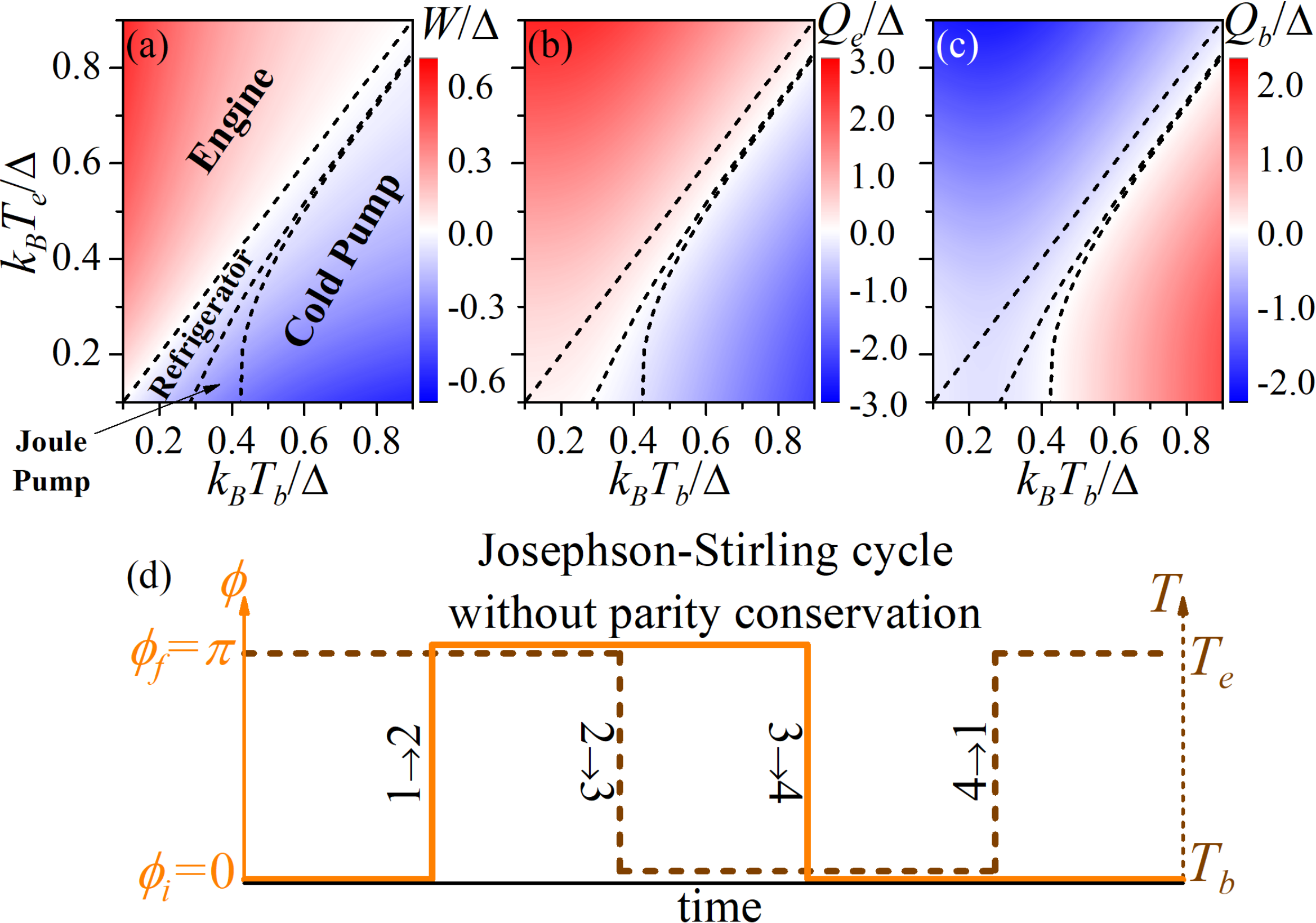}}\quad\quad
\subfloat[Josephson-Stirling cycle with parity constraints]{\includegraphics[width=8.5cm]{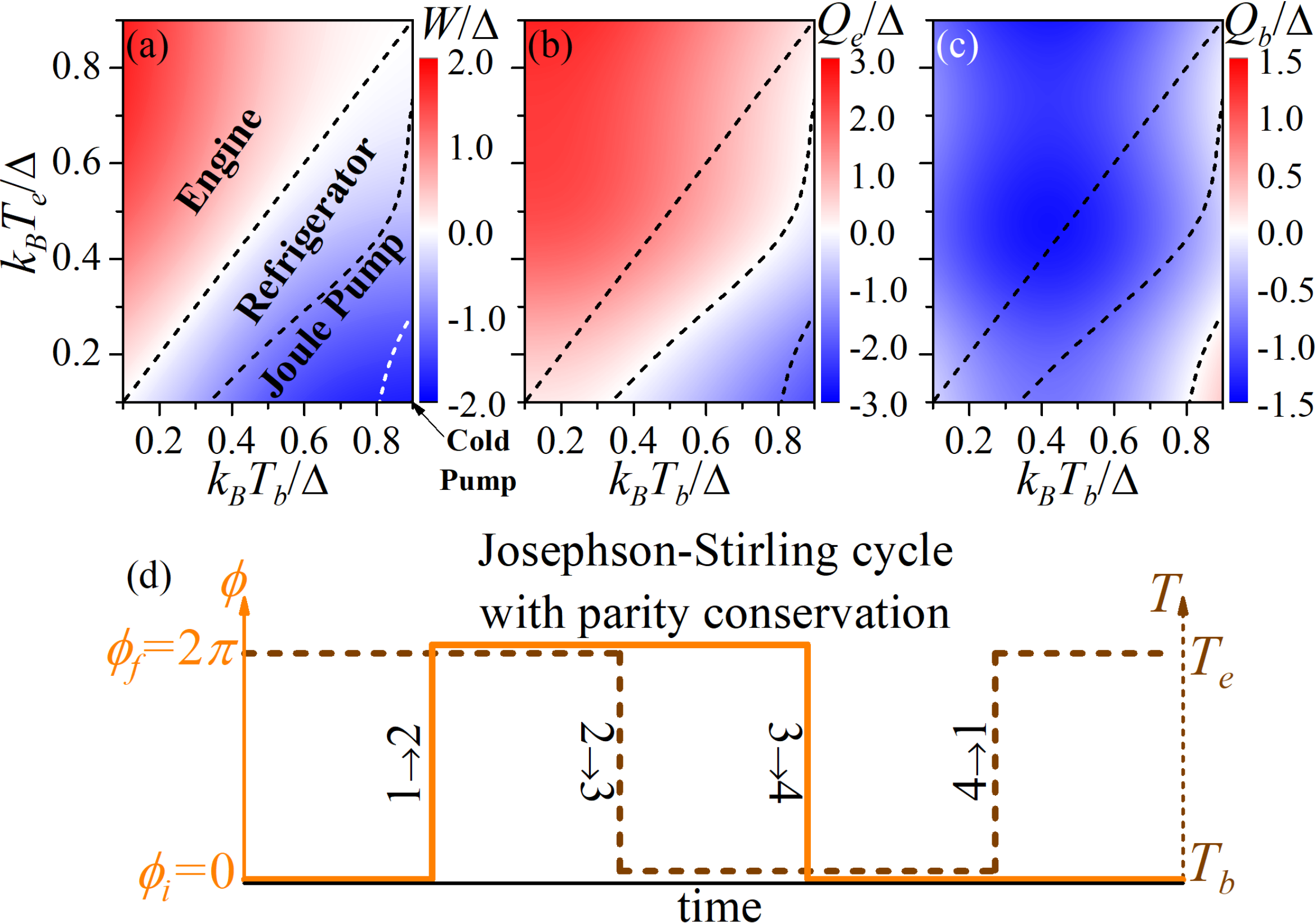}}\\\quad\\
\subfloat['Shifted' Josephson-Stirling cycle without parity constraints]{\includegraphics[width=8.5cm]{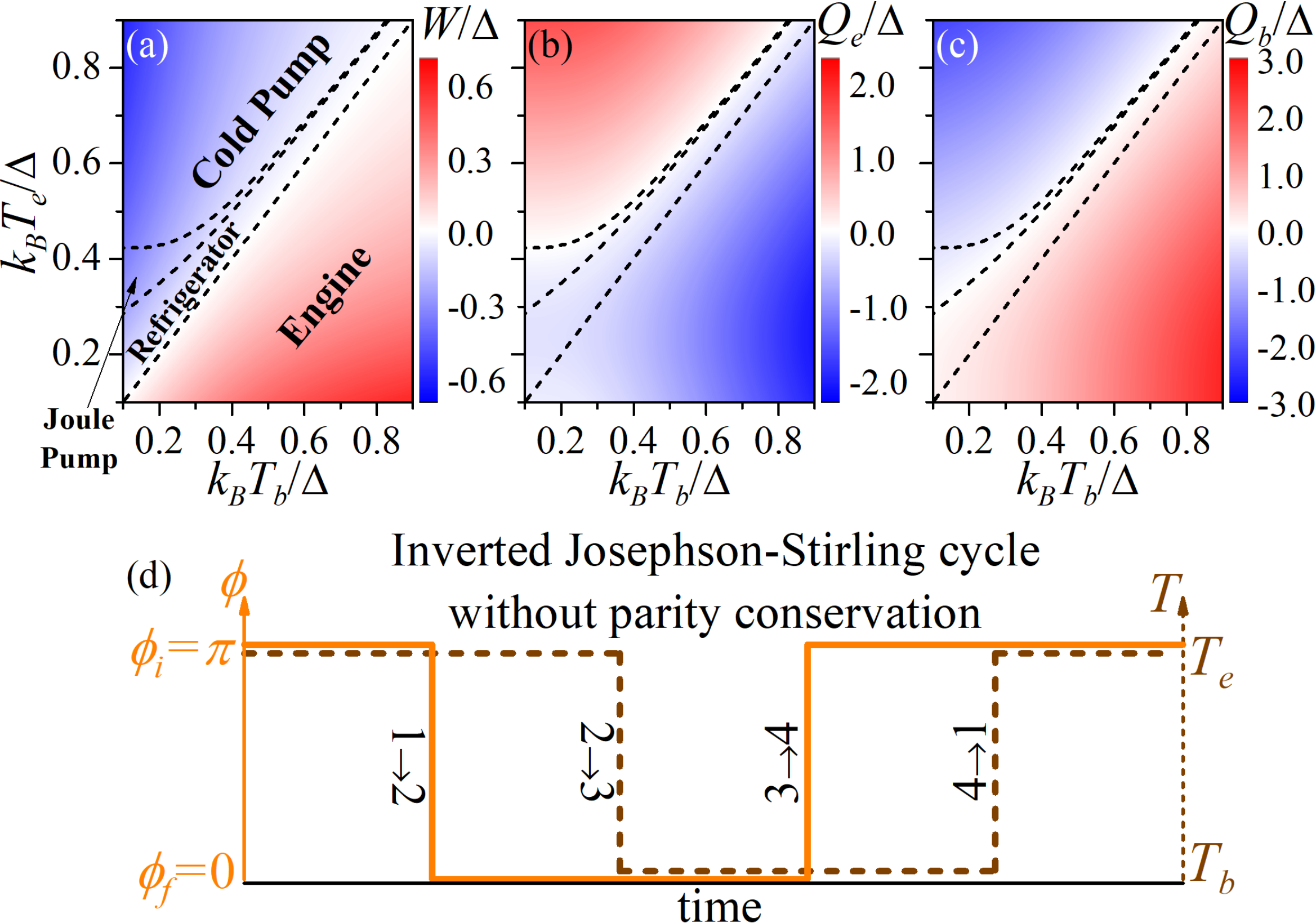}}\quad\quad
\subfloat['Shifted' Josephson-Stirling cycle with parity constraints]{\includegraphics[width=8.5cm]{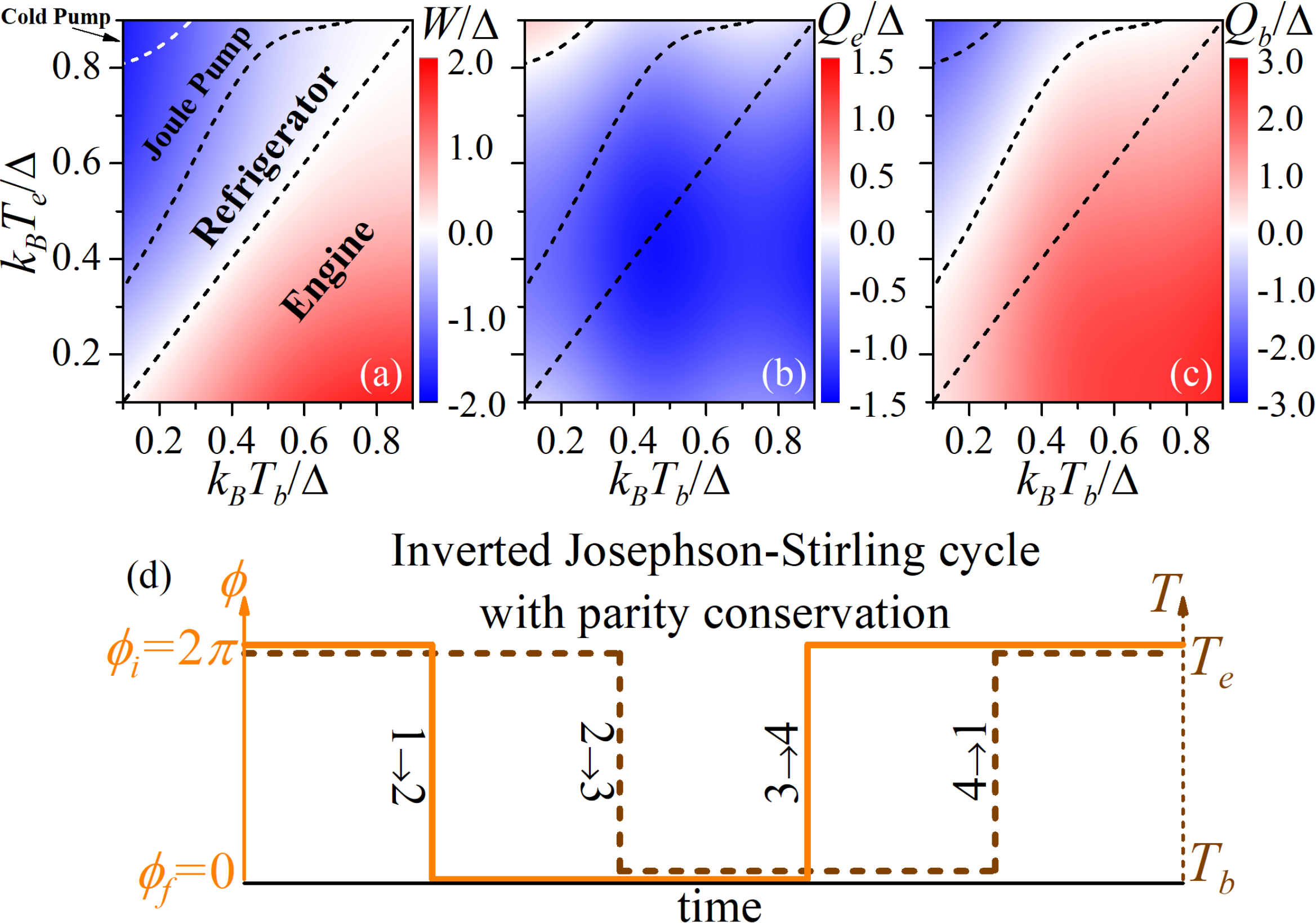}}
\caption{Different Josephson-Stirling cycles: (A) non-parity-conserving cycle, (B) parity-conserving cycle, (C) non-parity-conserving 'shifted' cycle, (D) parity-conserving 'shifted' cycle. In each subfigure, the panels show the (a) total work $W$ and heat exchanges (b) $Q_\mathrm{e}$ and (c) $Q_\mathrm{b}$ as functions of the reservoir temperatures $T_\mathrm{e}$ and $T_\mathrm{b}$. The different operating modes of each cycle are indicated in panels~(a). Panels~(d) show a scheme of the cycle considered. In all subfigures, $E_\mathrm{S}=0.165\Delta$.}\label{fig:StirlingNP}
\end{figure}

\section{Josephson-Stirling cycles}\label{App:PlotsStirling}
Here, we wish to discuss the flexibility of the topological engine to operate in different regimes. Figure~4 in the main text shows the different operating modes of a Josephson-Stirling cycle without [Figs.~4(a,b)] and with parity constraints [Figs.~4(c,d)], changing only the temperatures $T_\mathrm{e}$ and $T_\mathrm{b}$. In particular, we have presented the total work $W$ as well as the efficiency $\eta$ in the engine mode and the COP in the refrigerator mode. To identify these different modes for a given combination of temperatures $T_\mathrm{e}$ and $T_\mathrm{b}$, one has to compare $W$ with the heat exchanges $Q_\mathrm{e}$ and $Q_\mathrm{b}$.

For completeness, we present $W$, $Q_\mathrm{e}$ and $Q_\mathrm{b}$ for the Josephson-Stirling cycle without parity constraints and with $\phi_\mathrm{i}=0$ and $\phi_\mathrm{f}=\pi$ in Supplementary Fig.~\ref{fig:StirlingNP}(A) [compare to Figs.~4(a,b) in the main text]. Similarly, Supplementary Fig.~\ref{fig:StirlingNP}(B) shows $W$, $Q_\mathrm{e}$ and $Q_\mathrm{b}$ for the Josephson-Stirling cycle with parity constraints and with $\phi_\mathrm{i}=0$ and $\phi_\mathrm{f}=2\pi$ [compare to Figs.~4(c,d) in the main text]. In addition, we also present $W$, $Q_\mathrm{e}$ and $Q_\mathrm{b}$ for the above Josephson-Stirling cycles if $\phi_\mathrm{i}$ and $\phi_\mathrm{f}$ are exchanged. Supplementary Fig.~\ref{fig:StirlingNP}(C) shows the cycle without parity constraints and with $\phi_\mathrm{i}=\pi$ and $\phi_\mathrm{f}=0$, while Supplementary Fig.~\ref{fig:StirlingNP}(D) shows the cycle with parity constraints, $\phi_\mathrm{i}=2\pi$ and $\phi_\mathrm{f}=0$.

\section{Fermion parity conservation and thermodynamic cycles}\label{App:QP}

Having discussed schemes to determine the work per cycle via measuring the current-phase relation, we now turn to another experimental challenge: An experimental realization of our proposal to test the $4\pi$ periodicity of topological Josephson junctions requires that fermion parity is preserved.

In order to achieve this, one has to reduce quasiparticle poisoning, similar to other experimental proposals to directly observe a $4\pi$ periodicity in the junction properties. This requirement then precludes galvanic channels of heat transfer because large quasiparticle poisoning rates due to a direct galvanic contact to large thermodynamic reservoirs are expected to prevent conservation of the fermion parity in this case. For our scheme to work, however, a good thermal coupling of the QSH-based junction to superconducting reservoirs is not required, but rather a good thermal coupling to heat baths. Such a coupling does not necessarily have to be facilitated with galvanic coupling to baths. Radiative heating~\cite{Pendry1999:JPCM} could, for example, be a promising alternative to control the temperature of the QSH edges without the need for galvanic coupling to baths. In this scheme, the superconducting ring provides proximity-induced superconductivity, but operates at temperatures $T\ll T_c$ so that quasiparticle poisoning is reduced. The manipulation of the electronic temperature in the QSH edges arises from coupling to another thermal bath via radiative coupling. Importantly, radiative coupling does not necessarily require the thermal bath to be in direct galvanic contact with the thermal reservoir. Hence, we expect our proposed setup to not suffer from additional quasiparticle poisoning if non-galvanic thermal couplings are employed.

For the fermion parity to be preserved during one thermodynamic cycle, the quasiparticle poisoning time should be larger than the duration of the Josephson-Stirling cycle, which consists of two sweeps of the phase $\phi$ and two sweeps of the temperature of the QSH edge states.

\subsection{Temperature timescale}
The timescale of the temperature modulation is bounded from below by the response times for the QSH edge states to thermally equilibrate with the reservoirs. This response time can be roughly approximated as
\begin{equation}\label{eq:teq}
\tau_{eq}\approx C(\phi_0,T_\mathrm{e})/\kappa_{th}(T_\mathrm{e}),
\end{equation}
where $C(\phi_0,T_\mathrm{e})$ and $\kappa_{th}(T_\mathrm{e})$ are the heat capacity of the QSH edge states and the thermal conductance of the QSH-based Josephson junction with the thermal reservoir, respectively. Here, $\phi_0$ denotes the fixed superconducting phase difference at which the temperature is changed, that is, $\phi_0=0$ or $\phi_0=2\pi$ for the parity conserving case. 

For a rough estimate, we assume that the thermal conductance of a QSH edge is given by
\begin{equation}\label{eq:thCond}
\kappa_{th}(T)=\frac{\pi^2}{6}g_0\left(\frac{k_\mathrm{B}}{e}\right)^2T,
\end{equation}
where $g_0=2e^2/h$ is the quantum conductance of a single edge channel. This ideal estimate provides an order of magnitude for the temperature equilibration timescale, but non-ideal factors could further limit the thermal transfer efficiency. If we take the total heat capacity described in Supplementary Notes~\ref{App:TDproc} above, we find $\tau_{eq}\approx0.1$ ns as a rough estimate for the timescale of each of the isophasic temperature changes between $k_\mathrm{B}T_\mathrm{b}=0.1\Delta$ and $k_\mathrm{B}T_\mathrm{b}=0.4\Delta$ [$\tau_{eq}(\phi=0,T_\mathrm{e})\approx98.9$ ps and $\tau_{eq}(\phi=2\pi,T_\mathrm{b})\approx106.0$ ps].

\subsection{Quasiparticle poisoning rate}
As mentioned above, non-galvanic channels of heat transfer to the QSH system are preferable if fermion parity has to be preserved during the thermodynamic cycle. This precludes galvanic coupling of the QSH-based junction to thermodynamic reservoirs and we envision other channels of heat transfer, such as radiative heating~\cite{Pendry1999:JPCM}. In this case, additional poisoning from reservoirs is avoided and the quasiparticle poisoning rate should be comparable to those in microwave measurement setups of the fractional Josephson effect~\cite{Lutchyn2010:PRL,Chiu2019:PRB}.

While the quasiparticle poisoning timescales can be very large in conventional Josephson devices, where parity lifetimes of up to 1 minute have been reported in NbTiN~\cite{vanWoerkom2015:NP}, these timescales are not-well known for HgTe-based topological Josephson junctions. Following Ref.~\cite{Rainis2012:PRB}, the quasiparticle poisoning rate in topological superconductors ranges from $\Gamma_{qp}=0.1$ MHz to $\Gamma_{qp}=100$ MHz depending on the system details. The quasiparticle poisoning timescale is then given by $\tau_{qp}=1/\Gamma_{qp}$, which with the above estimates ranges from $\tau_{qp}=10$ ns to $\tau_{qp}=10$ $\mu$s and is typically estimated to be of the order of 1 $\mu$s~\cite{Virtanen2013:PRB,Frombach2020:PRB}.

Combining these estimates with our estimates for the thermal response time $\tau_{eq}\approx0.1$ ns (for each temperature change), one can see that the condition $\tau_{eq}\ll\tau_{qp}$ is easily satisfied in this case. At the same time, experimental realizations in qubit and SQUID technology show that modulation times of the order of 1 ns or even less can be achieved for controlling the phase $\phi$~\cite{Mueck2010:SST}. Judging from these simple estimates, it appears feasible to find regimes where the quasiparticle poisoning timescale is sufficiently large compared to the duration of a cycle, at least in setups without electronic coupling to reservoirs.

\begin{figure}[t]
\centering
% trim={<left> <lower> <right> <upper>}
\includegraphics*[trim=0 0 13cm 0,width=12cm]{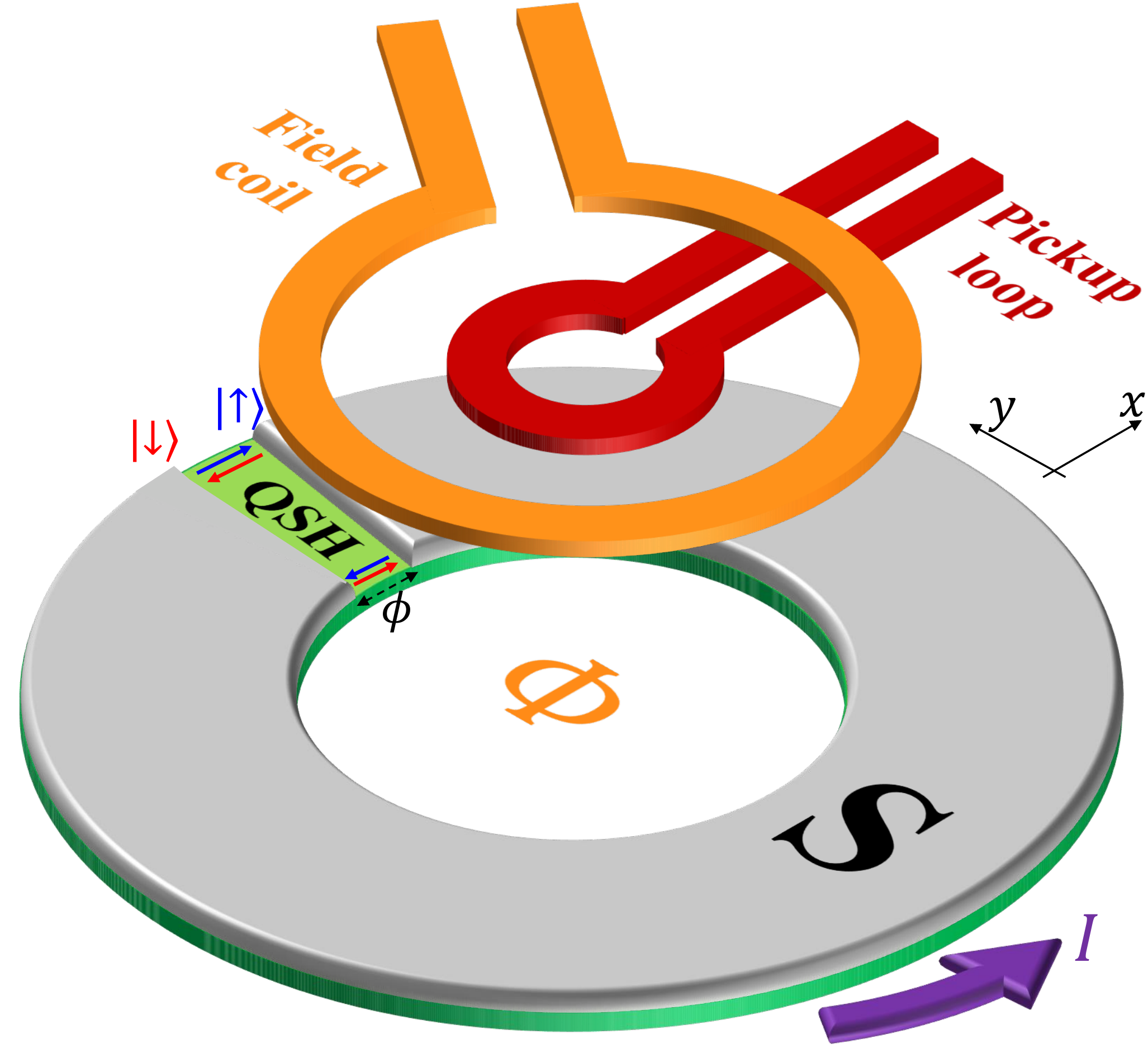}
\caption{Schematic of a measurement of the current-phase relation to determine the work produced during a Josephson-Stirling cycle. By driving a current through the field coil (orange) the flux $\Phi$ through the ring can be tuned. This in turn allows for control of the superconducting phase difference $\phi$ across the topological Josephson junction, which induces a supercurrent $I$ in the ring. The net supercurrent circulates in the ring, leading to a signal which is measured by the pickup loop of a SQUID sensor (red).}\label{fig:measurement}
\end{figure}

\section{Measuring the work per thermodynamic cycle}\label{App:Measurement}

Here, we discuss how the work per thermodynamic cycle can be experimentally determined. For a Josephson-Stirling cycle, work is done only during the isothermal processes $1\to2$ and $3\to4$ with 
\begin{equation}\label{eq:work}
W_{i\to f}=-\frac{\hbar}{2e}\int\limits_{\phi_\mathrm{i}}^{\phi_\mathrm{f}}\d\phi\; I(\phi,T)
\end{equation}
for each process, where $\phi_\mathrm{i}$ and $\phi_\mathrm{f}$ denote the corresponding initial and final phases. The total work per cycle is then given by the sum $W_{1\to2}+W_{3\to4}$. Hence, an experimental measurement of the total work requires a precise control of the phase bias $\phi$, modulated by the magnetic flux $\Phi$, as well as a good knowledge of the current $I$ flowing through the ring.

This can be achieved by using a scanning superconducting quantum interference device (SQUID) microscope to perform contactless measurements of the diamagnetic response and current-phase relation of the junction. The basic idea is shown in Supplementary Fig.~\ref{fig:measurement}, where the flux $\Phi$ through the ring can be tuned by driving a current through the field coil. This in turn allows for control of the superconducting phase difference $\phi$ across the Josephson junction, which induces a supercurrent $I$ in the ring. The net supercurrent $I$ circulates in the ring, leading to a signal which is measured by the pickup loop of a SQUID sensor. This scheme allows for direct low-noise measurements of the current-phase relation with a white-noise floor below 1 $\mu\Phi_0/\sqrt{Hz}$ and has been successfully employed for high-precision measurements of the current-phase relation in topological Josephson junctions based on HgTe and Nb superconductors~\cite{Sochnikov2015:PRL} as well as on InAs nanowires~\cite{Hart2019:PRB}. The current-phase relation measured in this way could then be used to compute the work integrals given by Eq.~(\ref{eq:work}).

\end{widetext}

%\begin{thebibliography}{9}
\bibliographystyle{apsrev4-2}
\bibliography{BibTopInsAndTopSup}
%\end{thebibliography}

\end{document}